\documentclass[%
 reprint,
nofootinbib,
 amsmath,amssymb,
 aps,
]{revtex4-2}
\usepackage{bm}

\usepackage{graphicx}
\usepackage{dcolumn}
\usepackage{bm}

    \usepackage{booktabs}
	\usepackage{amsmath}
	\usepackage{makeidx}
	\usepackage{amsfonts}
	\usepackage{subfigure}
	\usepackage{epsfig}
	\usepackage{footnote}
    \usepackage[bottom]{footmisc}
	\usepackage[colorlinks,hyperindex]{hyperref}
	\hypersetup
	{
		colorlinks,%
		citecolor=black,%
		linkcolor=black,%
		urlcolor=black,%
	}

\begin{document}

\preprint{APS/123-QED}

\title{Temporal networks with node-specific memory: unbiased inference of transition probabilities, relaxation times and structural breaks}

\author{Giulio Virginio Clemente}
\email{giulio.clemente@imtlucca.it}
\affiliation{%
 IMT School for Advanced Studies, Piazza San Francesco 19, 55100 Lucca (Italy)
}%

\author{Claudio J. Tessone}

\affiliation{
 Blockchain \& Distributed Ledger Technologies, UZH Blockchain Center, University of Zurich, Andreasstrasse 15, 8050 Zurich (Switzerland)
}%

\author{Diego Garlaschelli}
\affiliation{IMT School for Advanced Studies, Piazza S. Francesco 19, 55100 Lucca (Italy)}
\affiliation{Lorentz Institute for Theoretical Physics, Niels Bohrweg 2, 2333 CA Leiden (The Netherlands)}
\affiliation{INdAM-GNAMPA Istituto Nazionale di Alta Matematica (Italy)}%

\date{\today}

\begin{abstract}
One of the main challenges in the study of time-varying networks is the interplay of memory effects with structural heterogeneity. In particular, different nodes and dyads can have very different statistical properties in terms of both link formation and link persistence, leading to a superposition of typical timescales, sub-optimal parametrizations and substantial estimation biases.
Here we develop an unbiased maximum-entropy framework to study empirical network trajectories by controlling for the observed structural heterogeneity and local link persistence simultaneously. 
An exact mapping to a heterogeneous version of the one-dimensional Ising model leads to an analytic solution that rigorously disentangles the hidden variables that jointly determine both static and temporal properties. Additionally, model selection via likelihood maximization identifies the most parsimonious structural level (either global, node-specific or dyadic) accounting for memory effects.
As we illustrate on a real-world social network, this method enables an improved estimation of dyadic transition probabilities, relaxation times and structural breaks between dynamical regimes.
In the resulting picture, the graph follows a generalized configuration model with given degrees and given time-persisting degrees, undergoing transitions between empirically identifiable stationary regimes. 

\end{abstract}

\maketitle


\section{\label{sec:Introduction}Introduction}

Time-varying graphs have been studied intensively, due to the importance of evolving network structures for many systems and processes. 
Understanding the temporal variability of networks is indeed important, for different reasons, in several apparently unrelated fields. Examples include: (a) the study of communication in highly dynamic networks, e.g. broadcasting and routing in delay-tolerant networks \cite{vasilakos2016delay}; (b) the exploitation of passive mobility, e.g., the opportunistic use of transportation networks \cite{pentland2004daknet}; and (c) the analysis of complex real-world networks ranging from neuroscience and biology to transportation and social systems \cite{amaral2004virtual, barrat2008dynamical,caldarelli2007scale,barabasi2002evolution}, e.g., the characterization of the dynamic interaction patterns emerging in a social network \cite{newman2001structure,newman2004coauthorship,ghoshal2009random}.
As part of these research efforts, a number of important concepts have been identified, often only named, sometimes formally defined. Interestingly, it is becoming apparent that concepts introduced under different names and for distinct reasons are strongly related. For example, the concept of \emph{temporal distance}, formalized in \cite{xuan2003computing},  
is extremely close to the ones of {\em reachability time}~\cite{holme2005network}, {\em information latency}~
\cite{kossinets2008structure}, and {\em temporal proximity}~\cite{kostakos2009temporal}. Similarly, the concept of {\em journey} \cite{xuan2003computing} is virtually the same as {\em schedule-conforming path}~\cite{berman96vulnerability}, {\em time-respecting path}~\cite{holme2005network,kempe2000connectivity}, 
 and {\em temporal path}~\cite{chaintreau2007diameter,tang2010small}. 
 Hence, many of the notions discovered along different research avenues can be viewed as parts of the same conceptual universe, and several formal attempts arising in different specific concepts can be viewed as fragments of a potentially much larger formal description of this universe. 
 One of the most general  features common to all applications in these areas is the fact that the system's structure, the network topology, varies in time~\cite{HOLME}. Furthermore, the rate and/or degree of structural change is generally so large, compared with the other temporal scales of the problem, that changes cannot be considered anomalies but rather intrinsic features of the system's behaviour.
As the notion of (static) graph is the natural means for representing a static network  \cite{pastorsatorras2010complex}, the notion of dynamic (or time-varying, or evolving) graph is the natural means to represents these highly dynamic networks \cite{palla2007quantifying}. All the concepts and definitions advanced so far are based on or imply such a notion, as expressed even by the choices of names; e.g., Kempe et al. \cite{kempe2000connectivity} talk of a temporal network (G, $\lambda$) where $\lambda$ is a time-labeling of the edges, that associates punctual dates to represent dated interactions; Leskovec et al. \cite{leskovec2007graph} talk of graphs over time; Flocchini et al.  \cite{flocchini09exploration} and Tang et al. \cite{tang2010small}  independently employ the term time-varying graphs; Kostakos uses the term temporal graph \cite{kostakos2009temporal}.
In such dynamic settings, where the evolution of the networks is available, different ways to represent the data are proposed \cite{Masuda_Lambiotte}. 
Among those, a particular mention goes to the snapshots representation, where the time information is aggregated through non-overlapping fixed time windows, creating an ordinate sequence of static networks.
Even if this representation can be misleading \cite{Grindrod}, mainly for reasons linked to the time windows' choice, it is also true that choosing the right size for the window can bring different advantages \cite{temporal_aggregation}.
Snapshot representation results easy to interpret and gives a natural setup to readapt some models widely used for the static case.
Often the transposition of these models from the static to the dynamic case happens employing "memory" terms \cite{Leifeld}, usually of Markovian nature.  In this sense, for example, the Exponential random graph models (ERGMs) \cite{Wasserman_Pattison} are adapted to the dynamic case by adding different mechanisms to describe the temporal evolution. Following the first work where dynamic networks are described using a Markovian model \cite{Holland_Leinhardt}, Snijders \cite{Snijders} proposed an evolutionary mechanism that can be seen as the result of a continuous process,  where mini-steps describe the changing between two consecutive snapshots. Another way to readapt the ERGMs is by using a discrete mechanism, like in the case of the Temporal Exponential random graph models (TERGMs) \cite{Hanneke}.  In TERGMs there are no assumptions related to the sequence of creation/disruption of links between two consecutive snapshots. Links, in this approach, are generated simultaneously in each snapshot, following only the temporal order defined by the temporal network, with the primary purpose of replicating the constraints imposed on the model. It is in this last framework where we collocate our contribution.

Initially, we define a quantity that captures the temporal correlation between different snapshots: the persisting connectivity, an indicator for the stability of the connection between nodes.
Probably for its natural interpretation, link stability is a quantity popularly used in literature.
An implementation is operated in a model such as the "edge-markovian evolving graphs" \cite{Clementi}, where memory is studied observing the stability of links. Attentions on this mechanism are also placed in \cite{MAZZARISI202050}, which proposes a methodology to describe link persistence and the dynamic of some latent variables associated with each node that contribute to the link formation. Latent variables are also used in the work of Zhang et al. \cite{Zhang}, where the connection between pair of nodes is described by a continuous-time Markov process, with a characteristic rate depending on the various latent variables of the nodes.
Following some of the characteristics observed in the literature, we propose our models. We introduce different Maximum Entropy models, characterized by having the same functional form but being built in such a way that each of them can be seen as having different level of heterogeneity on the constraints used. The model works for stationary time networks.
We start with three models with no memory: the constraints used are, in order: a constraint for each link averaged in time, a constraint for each degree averaged in time, and for the last model, a constraint for the average degree averaged in time. 
At this point, we insert another kind of constraint to create a memory model, adding at the memory-less model, in which we constrain the average degrees, different kinds of specifications.
These specifications are related to the average in time of the persisting link, the average in time of the persisting degree, and the average in time of the average(over nodes) of degree persisting. 
After a suitable reparametrization, the resulting models show an analogy with the one-dimensional Ising model. Exploiting this analogy, we solved the model analytically, allowing us to use the maximum likelihood approach to estimate the model's parameters and compute exactly the expected value of some quantities of interest.

The proposed model is only applicable only to study temporal networks that present a stationary evolution.
We overcame this issue by proposing a structural break detection method that leverages these models to identify stationary periods. Thanks to this application, we are also able to capture changing points, often corresponding to exogenous events.

We apply all the models to study the MIT proximity Network (2004-2005) \cite{MIT_data}.
To judge the degree of heterogeneity, we use the Akaike Information Criterion (AIC) \cite{AIC}. First, we study only the memory-less models, observing that model with constraints on the degrees wins (according to AIC), then we move to the memory model, showing that the model with constraints on the degree and persisting degree wins the others, demonstrating that the system has a memory and that this memory can be explained by latent variable associated with each node. In the end, we characterized each node by exploiting the markovian process for the link formation in which it is involved.

\section{\label{sec:MAxEnt}Maximum-Entropy Probability for Complex Networks}

Before moving to the case of Temporal networks, we have to recall some theory that constitutes the base for the construction of our model for time-varying graphs.
Here we introduce the maximum entropy approach to complex networks as a powerful tool to model complex systems \cite{Max_Ent_book}.
Given a graph G, with a number of nodes N and some peculiar properties C(G), the maximum entropy approach describes a recipe to obtain the most unbiased probability distribution P(G) \cite{Jaynes} over the set of all possible graphs, of the same type, such that C(G) is on average reproduced.
To obtain such probability, we pass by a constrained entropy maximization \cite{Real_World_Networks}, that is to find the maximum of the Shannon-Gibbs entropy:
\begin{equation}
S(P(G)) = -\sum_{G} P(G)ln(P(G))
\label{eq:entropy_static}
\end{equation}

granting the constraints that the probability distribution has to reproduce on average

\begin{equation}
\langle C \rangle = \sum_G C(G) P(G) = C^*
\label{eq:constraints}
\end{equation}

Where G is a generic network in the ensemble, $C^*$ is the observed value of the constraint and $\langle \cdot \rangle$ indicates the expected value over the ensemble of networks.
The constrained maximization problem is solved by introducing a lagrange multiplier ($\theta$) for each constraint. The solution results in a probability distribution that turns out to have the same functional form of the Exponential Random Graphs Models (ERGMs) \cite{Park_Newman}:

\begin{equation}
P(G|\vec{\theta}) = \frac{e^{-H(G|\vec{\theta})}}{Z(\vec{\theta})}
\label{eq:ERG}
\end{equation}

Where $H(G|\vec{\theta}) = \sum_i \theta_i C_i$ is the graph Hamiltonian, and $Z(\vec{\theta}) = \sum_G e^{-H(G|\vec{\theta})}$ is the normalization constant, also called partition function.

A critical assumption that is often made when we work with these models is the independence of link formation, such that we are allowed to write the probability for the formation of the entire graph as:
\begin{equation}
P(\mathbf{G})\equiv\prod_{i<j}p_{ij}^{a_{ij}}(1-p_{ij})^{1-a_{ij}}
\label{eq:Pstatic}
\end{equation}
where $a_{ij}$ is the $i,j$ entry of the adjacency matrix of the graph $\mathbf{G}$, and $p_{ij}$ represents the probability that $a_{ij}=1$ (nodes $i$ and $j$ are connected).

A relevant example for the purpose of this paper is given by the Binary Configuration model (BCM), where the resulting probability distribution is obtained after constrained the degree sequence, such that it ammits an Hamiltonian representation:

\begin{equation}
H(\mathbf{G}|\vec{\theta})\equiv\sum_{i=1}^N \theta_i k_i
\label{eq:Hstatic}
\end{equation}
Where $k_i$ is the degree of node \textit{i}.
For this problem, the $p_{ij}$ is \cite{Garlaschelli_loffredo_wtw,likelihood_Garlaschelli_loffredo,Max_Ent_book} :
\begin{equation}
p_{ij}\equiv \frac{x_i x_j}{1+x_i x_j}.
\label{eq:pstatic}
\end{equation}

where $x_i\equiv e^{-\theta_i}$ for each node $i$ \cite{likelihood_Garlaschelli_loffredo,Max_Ent_book}.
$x_i$ is often called the 'fitness' of node \textit{i} and can be interpreted as the propensity of node \textit{i} to link with other nodes.

\subsection{Maximum likelihood principle}
When the model is applied to a real-world graph $\mathbf{G}^*$, one needs to introduce a criterion to select the values of the parameters $\vec{x}$. 
The maximum likelihood principle requires that the parameters are set to the particular value that maximises the likelihood of the data, given the model \cite{likelihood_Garlaschelli_loffredo}.
In other words, one should look for the value $\vec{x}^*$ that maximizes the likelihood function $P(\mathbf{G}^*|\vec{x})$.
It has been shown that, for the model considered here, the value $\vec{x}^*$ can be found as the solution of the following $N$ coupled equations \cite{likelihood_Garlaschelli_loffredo,Max_Ent_book}:
\begin{equation}
\langle k_i\rangle=\sum_{j\ne i}\frac{x_i x_j}{1+x_i x_j}=k^*_i\quad\forall i.
\label{eq:kconstraint}
\end{equation}
where $k_i^*\equiv k_i(\mathbf{G}^*)$ is the empirical degree of node $i$ in the real-world network $\mathbf{G}^*$.
In other words, the maximum likelihood principle shows that one needs to set 
$\vec{x}$ to the particular value $\vec{x}^*$ (or equivalently, to set $\vec{\theta}$ to the particular value $\vec{\theta}^*$) ensuring that each expected degree $\langle k_i\rangle=\sum_{j\ne i}p_{ij}$ matches the corresponding observed degree $k_i^*$.

This model reproduces the empirical topology of several real-world networks \cite{Max_Ent_book}. Moreover, in the particular case of the World Trade Web (WTW), i.e. the network of international trade among world countries \cite{Garlaschelli_loffredo_wtw}, it has been shown that the fitness $x_i^*$ can be associated with an empirical property, namely the Gross Domestic Product (GDP) of country $i$ \cite{Garlaschelli_loffredo_wtw, likelihood_Garlaschelli_loffredo}.

Given its theoretical and practical advantages, in what follows we will use the model specified by eq. \eqref{eq:pstatic} as a starting point for the extension of the formalism.

In general, if we require that a network model generates a desired set $\vec{C}\equiv\{C_1,\dots,C_K\}$ of topological properties, the maximization of \eqref{eq:entropy_static} leads to eq. \eqref{eq:ERG} where ${H(\mathbf{G})=\sum_{k=1}^K \theta_k C_k}$ \cite{Max_Ent_book,Park_Newman}.
If we apply the maximum likelihood principle, then we find that the value $\vec{\theta}^*$  maximizing the likelihood of the model \cite{likelihood_Garlaschelli_loffredo} is also the value ensuring that the expected value $\langle C_k\rangle$ of each property $C_k$ matches the corresponding empirical value $C_k^*\equiv C_k(\mathbf{G}^*)$.
Thus $\vec{\theta}^*$ can be found as the solution of the $K$ coupled equations $\langle C_k\rangle=C_k^*$ for all $k$.

\section{From static to Time-varying graphs}

There are different ways to move from a model for static graphs to one for time-varying graphs, mainly depending on the representation and on the assumption made on the system \cite{Masuda_Lambiotte}.
Our purpose is to create a model that generates a sequence of graphs with fixed number of nodes and with properties that can fluctuate in time but are on average constant.

\subsection{Extensions to time-varying graphs}
In this section, we discuss possible ways to create a model in order to generate a sequence 
\begin{equation}
{ \mathcal{G}}\equiv \{\mathbf{G}_1,\dots,\mathbf{G}_T\}
\end{equation}
of T temporal snapshots of a network, where we denote $\mathcal{G}$ as the `graph trajectory'. \\
For simplicity, we assume that the number $N$ of nodes is constant throughout the evolution of the graph, and that at each node can be attributed some quantity $x_i$ that is the analog of the fitness in eq.(\ref{eq:kconstraint}).

The simplest option is that of regarding the observed graph trajectory $\mathcal{G}$ as a sequence of independent realizations of the same process, the latter being still specified by the static probabilities $p_{ij}$ where $\vec{x}$ is time-independent.
The above specification amounts to assume that $\vec{x}$ is fixed and that the probability of the entire graph trajectory factorizes over all time steps as
\begin{equation}
\mathcal{P}(\mathcal{G}|\vec{x})=
\prod_{t=1}^TP(\mathbf{G}_t|\vec{x})
=\prod_{t=1}^T \prod_{i<j}p_{ij}^{a_{ij}(t)}(1-p_{ij})^{1-a_{ij}(t)}
\label{eq:Pfactor}
\end{equation}
Clearly, this means that we are not introducing any explicit statistical dependence among different temporal snapshots of the network.
So, while the dependence of all snapshots on the same fitness vector $\vec{x}$ implies that the graph trajectory will have a certain degree of stationarity (especially for the pairs of nodes with $p_{ij}$ close to either $0$ or $1$), the conditional independence (given $\vec{x}$) of different shapshots implies that there are no temporal correlations among such snapshots. 

\subsection{Dynamic fitness, no temporal dependencies}
Another possibility is to introduce the temporal dynamics via a time-dependence of $\vec{x}$, i.e. by replacing $x_i$ with $x_i(t)$ and consequently $p_{ij}$ with $p_{ij}(t)$ (while keeping the same functional form of the probability). In such a way, while different snapshots of the system are described by different probabilities, they are still statistically independent on each other.
This means that the probability of the graph trajectory $\mathcal{G}$ is still factorized:
\begin{eqnarray}
\mathcal{P}(\mathcal{G}|\vec{x})&=&
\prod_{t=1}^TP(\mathbf{G}_t|\vec{x}(t))\label{eq:Pfactor2}\\
&=&\prod_{t=1}^T \prod_{i<j}p_{ij}(t)^{a_{ij}(t)}[1-p_{ij}(t)]^{1-a_{ij}(t)}\nonumber
\end{eqnarray} 
where $\mathbf{x}$ is a $T\times N$ matrix with generic entry $x_i(t)$ in the $t$-th row and $i$-th column.
Note that assuming that the fitness is a dynamical variable $x_i(t)$ corresponds to the assumption that node fitness and network topology change over comparable time scales, while assuming that $x_i$ is a fixed variable corresponds to the assumption that the fitness of nodes changes much more slowly than the topology of the network. A notable contribution in this direction is the one of Sarkar and Moore \cite{DSNA}, where the authors introduced a latent distance model with Markovian property.

\subsection{Static fitness, temporal dependencies}
A third and final possibility is that of assuming that different snapshots are statistically dependent. This is a more fundamental change of the model, because it no longer implies that the different snapshots can be generated by a single function $p_{ij}$. 
In the simplest case, we can assume that $\vec{x}$ is time-independent as in eq. \eqref{eq:Pfactor}. However, now the probability of the graph trajectory no longer factorizes over time steps:
\begin{equation}
\mathcal{P}(\mathcal{G}|\vec{x})\ne
\prod_{t=1}^TP(\mathbf{G}_t|\vec{x})
\label{eq:Pnonfactor}
\end{equation}

important contribution in this direction is defined by the work of Hanneke \cite{Hanneke} or Zhang et al. \cite{Zhang}, where in both cases, the structural dynamic is assumed to have a Markovian nature while the parameters are kept constant.
Interesting is, as reported in the supplementary information, how the TERMs, introduce by Hanneke, can be seen as the result of a properly defined Entropy Maximization problem with suitable constraints.\\
Alternatively, we may assume that $\vec{x}$ is time-dependent as in  eq. \eqref{eq:Pfactor2}, as made in \cite{MAZZARISI202050},  but again the probability will not factorize. However, in this paper, we will focus on the case where $\vec{x}$ is fixed.
The structure of the temporal dependency among different snapshots must be specified explicitly, and will in general imply that the connection probability at time $t$ depends not only on $t$, but also on some $t'<t$.
Despite this, we want to keep the main property of the fitness model, i.e. the fact that the connection between nodes $i$ and $j$ only depends on properties of these two nodes. 
As a consequence, while $\mathcal{P}(\mathcal{G}|\vec{x})$ will not factorize over time steps, it will still factorize over pairs of vertices.
Introducing a solvable model of this scenario is our main goal.

\subsection{Measures of time correlation:  \emph{persisting connectivity}}
Before we provide an explicit model with time dependence, we introduce some useful definitions to capture time correlations.
In models with no memory, such as those defined by eqs.~\eqref{eq:Pfactor} and ~\eqref{eq:Pfactor2}, the conditional (given $\vec{x}$ or $\mathbf{x}$) independence of different snapshots implies
\begin{equation}
\langle a_{ij}(t)a_{ij}(t+\tau)\rangle_{nm}=
\langle a_{ij}(t)\rangle_{nm}\langle a_{ij}(t+\tau)\rangle_{nm}\quad\forall t,\tau>0
\label{eq:Adynamic}
\end{equation}
where the subscript \emph{nm} stands for `no memory'.

The above observation suggests that a useful quantity to introduce at the level of nodes is the \emph{persisting degree} $h_i(t,\tau)$, that we define as the part of the degree $k_i(t)$ that also persists at time $t+\tau$ ( or sum of links that persists $h_{ij}(t,\tau)$ ):
\begin{equation}
h_{i}(t,\tau)\equiv\sum_{j\ne i}a_{ij}(t)a_{ij}(t+\tau) = \sum_{j\ne i}h_{ij}(t,\tau).
\label{eq:hdef}
\end{equation}
The persisting degree is computed by summing the quantity $a_{ij}(t)a_{ij}(t+\tau)$, to which we will refer as \textit{persisting link} (a similar quantity is employed in \cite{Clementi,MAZZARISI202050,Zhang}).

Note that the value of $a_{ij}(t')$ at intermediate times ($t<t'<\tau$) has no effect on the above definition.
From eq. \eqref{eq:Adynamic}, the expected value of $h_i(t,\tau)$ under a model with no memory is always 
\begin{equation}
\langle h_{i}(t,\tau)\rangle_{nm}=
\sum_{j\ne i}\langle a_{ij}(t)\rangle_{nm}\langle a_{ij}(t+\tau)\rangle_{nm}\quad\forall t,\tau>0
\label{eq:hnm}
\end{equation}

Comparing the observed and expected (under a model with no memory) value of $h_i(t,\tau)$ is a useful criterion to measure time correlations in an observed graph trajectory. 
If we average over time, the quantity
\begin{equation}
A_i(\tau)\equiv \frac{1}{T}\sum_{i=1}^T\big[h_i(t,\tau)-\langle h_i(t,\tau)\rangle_{nm}\big]
\label{eq:A}
\end{equation}
 is a node-specific autocovariance function, which allows us to measure the time scale of the decay of memory for each node. 
If the measured autocovariance is zero for all nodes, then there is no need to introduce a model with memory, since a model described by eq. \eqref{eq:Pfactor} or \eqref{eq:Pfactor2} will be enough. 
By contrast, non-zero autocovariance can only be modelled using time dependencies of some form, thus leading to eq. \eqref{eq:Pnonfactor}.

Note that, in the static case defined by eq.~\eqref{eq:Pfactor}, 
\begin{eqnarray}
\langle a_{ij}(t)a_{ij}(t+\tau)\rangle_{nm}&=&p_{ij}^2\quad\forall t,\tau>0.
\label{eq:Astatic}\\
\langle h_{i}(t,\tau)\rangle_{nm}&=&
\sum_{j\ne i}p^2_{ij}\quad\forall t,\tau>0
\label{eq:hstatic}
\end{eqnarray}
These expressions will be useful in the following.

\section{A solvable model with memory}
We now consider an explicit, solvable model with time correlations. For the sake of simplicity, we assume that $\vec{x}$ is fixed, as we already mentioned.\\
Our model is built by exploiting the Maximum Entropy formalism described above, and re-adapting it to describe ensemble of network trajectories (see Appendix). In doing so we define the entropy for the network trajectories that in general reads: 
\begin{equation}
\mathcal{S}(\mathcal{P}(\mathcal{G}))\equiv-\sum_{\mathcal{G}}
\mathcal{P}(\mathcal{G})
\ln \mathcal{P}(\mathcal{G})
\label{eq:entropy_temporal}
\end{equation}

Hence, by looking for $P(\mathcal{G})$ that maximizes (\ref{eq:entropy_temporal}), such that some properties observed on the trajectory $C_i(\mathcal{G})$ are on average replicated, we get the solution:
\begin{equation}
    P(\mathcal{G}|\vec{\theta}) = \frac{e^{-\sum_i \theta_i C_i(\mathcal{G})}}{Z(\vec{\theta})}
    \label{TERGM_general_main}
\end{equation}
Where
\begin{equation}
    Z(\vec{\theta}) = \sum_{\mathcal{G}} e^{-\sum_i \theta_i C_i(\mathcal{G})}
\end{equation}
Is the partition function.\\

By employing the persisting connectivity, we define a particular kind of Maximum Entropy model for temporal network that can be seen as a natural generalization of the Binary Configuration Model.

\subsection{Preliminary extension: the memoryless model}
As a preliminary step, we first consider three (memory-less) models defined by using three different kind of constraints.
The constraints are presented in order to reproduce a decreasing amount of heterogeneity among nodes.\\
The first model reproduces exactly the average of each link, where the correspondent Hamiltonian is:
\begin{eqnarray}
\mathcal{H}_{1,nm}(\mathcal{G}) \equiv\frac{1}{T}\sum_{t=1}^T H_1(\mathbf{G}_t)
&=&  \nonumber \\ 
= \frac{1}{T}\sum_{t=1}^T \sum_{j<i} \alpha_{ij} a_{ij}(t) &=& \sum_{j<i} \alpha_{ij} \overline{a_{ij}}
\label{eq:Hbar_1}
\end{eqnarray}
This first specification leads to a nontrivial model because contrary to the static case, here we constrain the links averaged\footnote{The values constrained can assume values $\in [0,1]$, and not only 0 or 1 as in the static binary case.} in time, making the model nondegenerative: the lagrange multipliers can assume values different from 0 or $\infty$ (that is what would happen in the static case).

The second kind of constraint is similar to the one for the static BCM, where the difference is given by the fact that the degrees are averaged in time. This results in the following Hamiltonian:

\begin{eqnarray}
\mathcal{H}_{2,nm}(\mathcal{G})
&\equiv& \frac{1}{T}\sum_{t=1}^T H_2(\mathbf{G}_t)
= \nonumber \\
&=&\frac{1}{T}\sum_{t=1}^T\sum_{i=1}^N \alpha_i {k}_i(t)
=\sum_{i=1}^N \alpha_i \bar{k}_i
\label{eq:Hbar_2}
\end{eqnarray}

The last kind of memory-less model is built taking as constraint the average over time and over nodes of the degree.
This turns out to have a structure analogous to the Erdős–Rényi model, having the following Hamiltonian:

\begin{eqnarray}
\mathcal{H}_{3,nm}(\mathcal{G})
&\equiv& \frac{1}{T} \sum_{t=1}^T H_3(\mathbf{G}_t)= \nonumber \\
&=&\frac{1}{T N}\sum_{t=1}^T\sum_{i=1}^N \alpha {k}_i(t) 
=\frac{\alpha}{N} \sum_{i=1}^N  \bar{k}_i
\label{eq:Hbar_3}
\end{eqnarray}

In other words, these are models that maximizes the entropy of the entire ensemble of graph trajectories, defined in eq. (\ref{eq:entropy}) under constraints defined by each specification.

If we apply the maximum likelihood principle to each of these models, i.e. if we look for the specific value $\vec{\alpha}^*$ that maximizes the likelihood $\mathcal{P}(\mathcal{G}^*|\vec{\alpha})$ of an observed graph trajectory $\mathcal{G}^*$, we find that for the three specification, the probability of a link in each snapshot between two nodes, is given by the following:

\begin{equation}
p_{ij} = \frac{x_{ij} }{1+x_{ij} }
\label{eq:p_ij_general}
\end{equation}

Where

\begin{equation}
x_{ij} = 
\begin{cases} 
x_{ij} = 
e^{-\alpha_{ij}/T} &\text{if} \quad  \mathcal{H}_{1,nm}(\mathcal{G})\\
x_i x_j = e^{-\alpha_i/T} \cdot e^{-\alpha_j/T} &\text{if} \quad \mathcal{H}_{2,nm}(\mathcal{G})\\
x = e^{-\alpha/TN} &\text{if} \quad \mathcal{H}_{3,nm}(\mathcal{G})
\end{cases}
\label{eq:x_ij}
\end{equation}
In Appendix we show how we find $\vec{\alpha}^*$ for the three specifications.

While this kind of models can replicate any of the three observed time-averaged quantities that we constraint, they predict no time correlations. Using eq. \eqref{eq:p_ij_general}, we find that for the three specifications, the expected value of the persisting degree is

\begin{equation}
\langle h_{i}(t,\tau)\rangle=
\sum_{j\ne i}\left(\frac{x_{ij}}{1+x_{ij}}\right)^2\quad\forall t,\tau>0.
\label{eq:hexp1}
\end{equation}

\begin{figure}[h!]
\includegraphics[width=7cm]{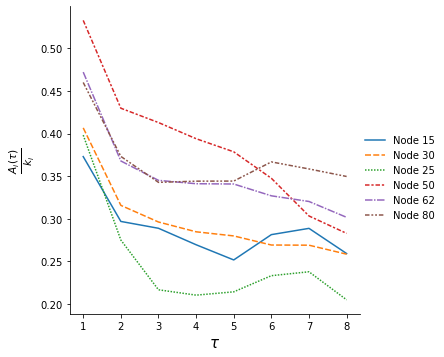}
\caption{In figure we have the plot of the normalized empirical autocovariance, eq.\eqref{eq:A}, as a function of $\tau$ for various nodes of the MIT proximity dataset: the outcome is a non-zero decreasing function.}
\label{real_autocorrelation}
\end{figure}

Comparing the above prediction with the observed value of $h_{i}(t,\tau)$ is a simple test of the adequacy of the models. 
In particular, the expected value of the node-specific autocovariance $A_i(\tau)$ defined in eq. \eqref{eq:A} is
\begin{equation}
\langle A_i(\tau)\rangle= \frac{1}{T}\sum_{i=1}^T\big[\langle h_i(t,\tau)\rangle-\langle h_i(t,\tau)\rangle_{nm}\big]=0
\end{equation}
The figure (\ref{real_autocorrelation}) displays the values of the normalized autocovariance function relative to the degree, representing various nodes from the dataset analyzed in the empirical application of the model.

From this plot we can note that the behaviour for the node-specific autocovariance is different for each node and moreover different from 0.  This first observation gives already an intuition about the heterogeneity in the memory associated with each nodes, and it raises the necessity of having a way to capture it.


\subsection{The full model}
Now we extend the above model by adding temporal correlations at different degrees of heterogeneity, in such a way that the expected persisting degree is non-trivial and the autocorrelation is non-zero. 
As made for the memory-less case, here we define three different constraints that are added to the Hamiltonian defined by eq. \eqref{eq:Hbar_2}, resulting in three different models.\\
The simplest possibility is that of constraint the one-lagged persisting link $\overline{a_{ij}(t)a_{ij}(t+1)}$.
The second possibility is to constraint the one-lagged persisting degree $\overline{h_{i}(t,1)}$, and lastly we can constraint the average over all nodes of the one-lagged persisting degree, $\overline{h(t,1)}$:

\begin{equation}
    \overline{h(t,1)} = \frac{1}{N} \sum_i^N \overline{h_{i}(t,1)}
    \label{ave_persisting}
\end{equation}

As a result, the empirical value of $h_{i}(t,1)$, as well as that of the one-lagged autocorrelation $A_i(1)$, will be replicated exactly for the first two models, while on average for the third one.

Therefore, the goal of the models becomes that of checking whether the one-step memory is enough in order to reproduce the quantities $h_{i}(t,\tau)$ and $A_i(\tau)$ for all $\tau$; and understand the level of heterogeneity in terms of memory within nodes.

In this way, the three Hamiltonians to add at eq.\eqref{eq:Hbar_2} are the following:
\begin{widetext}
\begin{equation}
\mathcal{H}_m(\mathcal{G}) = 
\begin{cases} 
 \mathcal{H}_{m,1}(\mathcal{G})=\frac{1}{T} \sum_{j<i} \sum_t^T\beta_{ij} a_{ij}(t) a_{ij}(t+1) 
 &\text{Average over time of }\\ &\text{ the persisting link} \\
 \mathcal{H}_{m,2}(\mathcal{G})=\frac{1}{T} \sum_{i}^N \sum_t^T \beta_i h_i(1,t) &\text{Average over time}\\ &\text{  of the persisting degree} \\
 \mathcal{H}_{m,3}(\mathcal{G})=\frac{1}{TN} \sum_{i}^N \sum_t^T \beta h_i(1,t) &\text{Average over nodes} \\ &\text{of the average persisting degree}
\end{cases}
\label{eq:x_ij_2}
\end{equation}
\end{widetext}

Where the subscription \textit{m} stands for memory.\\
In order to solve analitically these models, for simplicity, we assume periodic boundary conditions, i.e. we add a fictious time step $T+1$ such that
\begin{equation}
a_{ij}(T+1)\equiv a_{ij}(1)
\label{eq:periodic}
\end{equation}
The effects of this artefact are more and more irrelevant as $T$ grows.\\
We therefore define our models through the following extension of eq.\eqref{eq:Hbar_2}:
\begin{equation}
\mathcal{H}(\mathcal{G})\equiv
 \mathcal{H}_{2,nm}(\mathcal{G}) + \mathcal{H}_m(\mathcal{G}) 
\label{eq:Hkh}
\end{equation}

As we show in the Appendix, after a suitable reparametrization the above models can be mapped exactly to a superposition of non-interacting one-dimensional Ising models. 
The models can then be solved analytically, and their  solutions can be given in terms of a constant connection probability 
\begin{equation}
p_{ij}=\langle \overline{a_{ij}(t)}\rangle
\end{equation}
and of a `memory' function 
\begin{equation}
q_{ij}(\tau)=\langle \overline{a_{ij}(t)a_{ij}(t+\tau)}\rangle
\end{equation}
Explicitly,
\begin{eqnarray}
 &p_{ij}= \left(\frac{x_{i}x_{j} y_{ij} -1}{2\sqrt{4x_{i}x_{j} + (x_{i}x_{j} y_{ij} - 1)^2}} \right)\left( \frac{(\lambda_{ij}^+)^T - (\lambda_{ij}^-)^T}{(\lambda_{ij}^+)^T + (\lambda_{ij}^-)^T} \right)  +\frac{1}{2}& \nonumber \\
 \\
&q_{ij}(\tau)=p_{ij}^2+p_{ij}(1-p_{ij})\left(  \frac{(\lambda_{ij}^-)^\tau (\lambda_{ij}^+)^{T-\tau}  + (\lambda_{ij}^+)^\tau (\lambda_{ij}^-)^{T-\tau}    }{(\lambda_{ij}^+)^T + (\lambda_{ij}^-)^T} \right)& \nonumber\\
\end{eqnarray}
With 
\begin{eqnarray}
\lambda_{ij}^{\pm} = e^{J_{ij}} \cosh B_{ij}\pm\sqrt{e^{2J_{ij}}\sinh^2 B_{ij}+e^{-2J_{ij}}}
\end{eqnarray}
and 
\begin{eqnarray}
B_{ij}&\equiv& \frac{1}{2}\left[ \ln(x_{i})+ \ln(x_{j})+\ln(y_{ij})\right]\\
J_{ij}&\equiv& \frac{1}{4}\left[ \ln(y_{ij}) \right]
\end{eqnarray}

Here, $x_{i} = e^{-\alpha_i/T}$, while $y_{ij}$ assumes different expression depending on the level of heterogeneity constrained on the model.
Like $x_{ij}$ for the different memory-less models, we have:
\begin{equation}
y_{ij} = 
\begin{cases} 
y_{ij} = 
e^{-\beta_{ij}/T} &\text{Persisting link}\\
y_i y_j = e^{-\beta_i/T} \cdot e^{-\beta_j/T} &\text{Persisting degrees}\\
y = e^{-\beta/TN} &\text{Average Persisting degree}
\end{cases}
\end{equation}
The above expressions imply that the conditional probability that nodes $i$ and $j$ are connected at time $t+\tau$, given that they were connected at time $t$, is
\begin{eqnarray}
P\big[a_{ij}(t+\tau)=1 \big| a_{ij}(t)=1\big]=
\frac{q_{ij}(\tau)}{p_{ij}}\\
=
p_{ij}+(1-p_{ij})\left(  \frac{\lambda_-^\tau \lambda_+^{T-\tau}  + \lambda_+^\tau \lambda_-^{T-\tau}    }{\lambda_+^T + \lambda_-^T} \right)
\end{eqnarray}
This shows how the models can generate non-trivial temporal dependencies.

The expected node-specific (Normalized)  autocovariance function is 
\begin{eqnarray}
&\langle A_i(\tau)\rangle=\sum_{j\ne i} p_{ij}(1-p_{ij})\left(  \frac{(\lambda_{ij}^-)^\tau (\lambda_{ij}^+)^{T-\tau}  + (\lambda_{ij}^+)^\tau (\lambda_{ij}^-)^{T-\tau}    }{(\lambda_{ij}^+)^T + (\lambda_{ij}^-)^T} \right)& \nonumber \\
\end{eqnarray}

\section{Transition Matrices and Correlation Length}
Differently from the static and the memory-less cases, here, to generate the network trajectory is not enough $p_{ij}$, but instead, we need a combination $p_{ij}$ and $q_{ij}$ to create a coherent time-varying graph.\\
Specifically, given the assumption of independence in link formation, we need $N(N-1)/2$ transition matrices from which, given an initial configuration, we can generate the entire graph trajectory. To build these matrices, we use the quantities estimated in our models. For each link, the stochastic matrix will have the same functional form, despite the different values for the elements.\footnote{See Appendix for further details.}

\begin{equation}
P_{ij} = 
\begin{bmatrix}
\frac{q_{ij}}{p_{ij}} & \frac{p_{ij}-q_{ij}}{p_{ij}} \\

\frac{p_{ij}-q_{ij}}{1-p_{ij}}  &  \frac{1-2p_{ij}+q_{ij}} {1-p_{ij}}
\end{bmatrix}
\end{equation}

This matrix has the same form independently of the model specification chosen: the only difference is in the form assumed by $p_{ij}$ and $q_{ij}$.
After obtaining the transition matrices, we can utilize them to characterize the stationary distribution between pairs of nodes as well as the average time needed to reach this state. It is worth noting that, by construction, the largest eigenvalue of the stochastic matrix is equal to one, and its associated eigenvector is $(p_{ij}, 1-p_{ij})$ (refer to the Appendix for more details). This implies that the stationary connection probability corresponds precisely to the marginal connection probability $p_{ij}$. Additionally, the second largest eigenvalue $\mu_{ij}$ of the stochastic matrix equals
\begin{equation}
\mu_{ij} = \frac{q_{ij}-p_{ij}^2}{p_{ij}-p_{ij}^2}
\label{eigenvalue_2}
\end{equation}

and is related to the rapidity of convergence to the stationary distribution.
Indeed, given the mapping of the system to the one dimensional Ising model, $\mu_{ij}$ turns out to be related to the correlation length $\tau^{c}_{ij}$ (which here plays the role of a correlation time), as follows:
\begin{equation}
\mu_{ij} = \left(  \frac{\lambda_- \lambda_+^{T-1}  + \lambda_+ \lambda_-^{T-1}    }{\lambda_+^T + \lambda_-^T} \right)  = e^{-\frac{1}{\tau^{c}_{ij}}}.
\label{correlation_length_exp}
\end{equation}

This implies:

\begin{equation}
\tau^{c}_{ij} = \frac{1}{\ln \left( \frac{1}{\mu_{ij}}\right)}
\label{correlation_length}
\end{equation}

The correlation time $\tau^{c}_{ij}$ tells us how strong the memory is, i.e. how fast the correlation vanishes as a function of the time lag between two snapshots. Clearly, different pairs of nodes have different values of $\tau^{c}_{ij}$, so some pairs require more time to relax to their stationary connection probability $p_{ij}$.
In principle, one can also focus on the neighbourhood of each node $i$ and look at the largest second eigenvalue $\mu_i\equiv\textrm{max}_{j\ne i}\{\mu_{ij}\}$ among the $N-1$ pairs involving that node, which can be used as a node feature defining the maximum time $\tau^c_i\equiv\textrm{max}_{j\ne i}\{\tau^c_{ij}\}=-1/\ln\mu_i$ to wait before reaching the stationary distribution for the entire neighbourhood of node $i$. Therefore, even when studying a system that is already in equilibrium and assuming that our model accurately describes it, understanding the correlation times can provide insights into the anticipated duration for the system to return to equilibrium after experiencing a transient regime due to an applied perturbation.

\section{Structural Break Detection}

To apply the model to a real-world dataset, we must address an important issue: real-world temporal networks often exhibit changing points, which signify a shift in the statistical properties of the system and render the model introduced earlier inappropriate. Although this is a common occurrence, by carefully selecting an appropriate time scale and window, it is possible to identify segments where the statistical properties of the system can be treated as stationary. The problem of detecting structural breaks has been tackled using various techniques, ranging from model-based approaches \cite{Change_points_SBM,Peel_Clauset,CPDSBM} to model-free methods \cite{CP_graphon}.

In this section, we present a model-dependent methodology to address this task, which can be easily adapted to the different models introduced earlier.

Our approach utilizes a widely recognized technique in the field of detecting structural breaks called Binary Segmentation \cite{Scott_Knott}. This methodology relies on an appropriate cost function that measures the level of uniformity among observations. The selection of the correct cost function is crucial as it directly reflects the underlying mechanism we believe governs the temporal evolution and characterizes the structural breaks. In our case, we employ a cost function based on the Akaike Information Criterion (AIC) \cite{AIC}. By incorporating AIC, we not only consider the likelihood value but also take into account the number of parameters. The AIC is defined as follows:
\begin{equation}
    AIC_{model} = 2 K_{model} - 2 \textit{L}_{model}
\end{equation}

Where $K_{model}$ corresponds to the number of parameters of the model, and $\textit{L}_{model}$ to the likelihood, computed by fitting the data.
Hence, breaks are defined by solving the following problem:

\begin{equation}
t_{break} = \underset{t \in [1, \cdots, T-1]}{argmin}  \Delta AIC(t)
    \label{binseg}
\end{equation}

Where $\Delta AIC(t)$ is our cost funciton and, given a temporal network of length T, is defined in the following:

\begin{equation}
\Delta AIC(t) = AIC_{tot} - AIC_{diff}
\end{equation}
 
 Where 
 \begin{equation}
      AIC_{tot} = 2 K_{model} - 2 \textit{L}_{model}
 \end{equation}

While
\begin{equation}
    AIC_{diff}(t) =  4 k_{model} - 2 (\mathcal{L}(\mathcal{G}_{0 , t}) + \mathcal{L}(\mathcal{G}_{t , T}))
\end{equation}

$t_{break}$ is accepted only if $\Delta AIC(t_{break}) < 0$, otherwise we assume there are no breaks.\\
The underlying idea is to compare a model fitted to the entire temporal network, assuming a single set of parameters, with a model that considers parameter changes at a specific point ($t_{\text{break}}$). The latter model incorporates twice the number of parameters as the initial set, capturing the potential structural changes in the system.

Once a structural break is identified, we isolate the new segment $[0, t_{\text{break}}]$ and repeat the procedure on this updated temporal network. We continue this process until further division is no longer advantageous, and we designate the final break as $t_{\text{break final}}$.

At this stage, we remove the segment $[0, t_{\text{break final}}]$ and focus on the remaining segment $[t_{\text{break final}}, T]$. Consequently, we obtain a collection of segments that, according to the assumed model, can be considered as generated by models with distinct parameters while being approximately constant within each segment.

It is worth noting that the number of identified breaks implicitly influences our analysis. A higher number of breaks indicates poorer model performance, thereby favoring the selection of models with fewer structural breaks.

As we show in the application, whether the identified points correspond to actual events or genuine changes in the data-generating process depends on the degree to which the employed model accurately describes the data.

\section{Application to Social proximity network}

For our real-world application, we examine a proximity temporal network derived from an experiment conducted by the Massachusetts Institute of Technology (MIT). Specifically, during the academic year 2004-2005, the MIT Media Laboratory conducted a reality mining experiment \cite{MIT_data}. The experiment involved tracking ninety-two participants through their smartphones. Utilizing Bluetooth data, we obtained proximity measurements, which we interpret as connections between the individuals. By organizing the linkages on a daily basis and recording the timestamps of these proximity events, we constructed a daily empirical temporal network consisting of 244 undirected snapshots.

Before directly applying the models to the dataset, we conducted a preliminary analysis to assess the significance of the snapshot order. To address this question, we generated 1000 instances of the same temporal network, with each instance constructed by completely shuffling the temporal order of the snapshots. Subsequently, we computed the persisting degree for various values of $\tau$ on each sample.

\begin{figure}[h!]
\includegraphics[width=7cm]{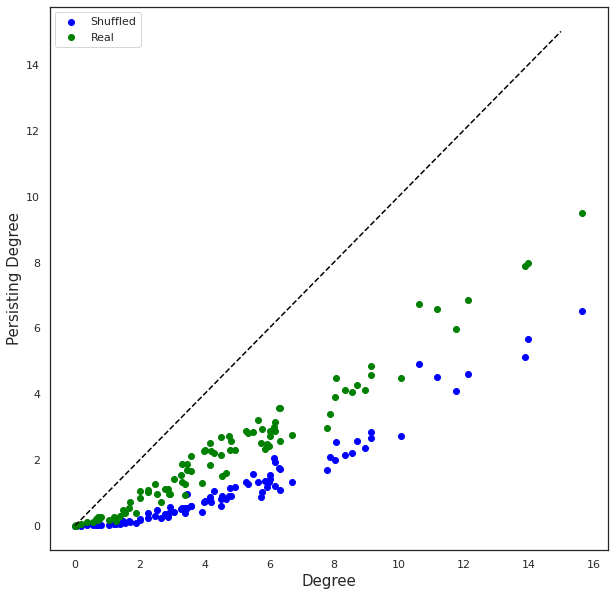}
\centering
\caption{Here we show the comparison between the persisting degree at $\tau=1$ computed on the real network and the average of the same quantity on the shuffling system.}
\label{autocorrelazione_null}
\end{figure}

Figure (\ref{autocorrelazione_null}) illustrates that the persisting degree, computed by averaging over the reshuffled temporal network, is lower than the one computed on the original instance. This finding indicates the presence of a non-trivial temporal ordering. To further investigate the significance of the difference between the persisting degree on the original network and its average value on the reshuffled system, we examine the behavior of each individual node.

For each node in the network, we calculate the z-score, which represents the number of standard deviations by which the randomized value of the persisting degree, obtained from the 1000 temporal randomizations, deviates from the persisting degree measured on the original network. This analysis is performed for three different values of $\tau$.

We observe that as $\tau$ increases, the number of nodes exhibiting a significant positive divergence in the persisting degree, compared to the value obtained from randomization, remains constant. This behavior is atypical of Markovian systems.

Specifically, for $\tau=1$, $\tau=2$, and $\tau=3$, approximately $97\%$ of the total nodes display a z-score greater than 1.96, indicating a significant positive difference.

This preliminary analysis suggests that the system deviates significantly from being Markovian and is not adequately described by the introduced model. However, it is important to note that this analysis does not consider the potential presence of structural breaks, which may conceal a behavior more closely aligned with our models. To address this, we employ the aforementioned structural break techniques to detect and account for changing points.

In our analysis, we first examined the three memory-less models to assess the level of heterogeneity associated with the link formation mechanism, before investigating memory effects. During the structural break detection process, we aim to identify a vector of $\mathcal{N}$ breaks:

\begin{equation}
    T_{v} = [t_{break}(0),..., t_{break}(\mathcal{N})]
\end{equation}

Once those breaks are defined, we can compute the value of the AIC for each bounded period. 

Where, for each period, AIC depends on the likelihood and on the number of parameters each model uses.
For the three memory-less cases, we have that 

\begin{equation}
K_{model} = 
\begin{cases} 
 N (N-1)/2& \text{for} \ \mathcal{H}_{1,nm}\\
 N &\text{for} \ \mathcal{H}_{2,nm}\\
 1 &\text{for} \ \mathcal{H}_{3,nm}\\
\end{cases}
\end{equation}

And the likelihood measures the ability to reproduce the average over time of the observed links.\\

To compare which of the models explain better the data, we compute the total AIC for the three models, defined as the summation of the AIC computed for each segment individuate:

\begin{equation}
    AIC_{total} = \sum_{t_b \in T_{v}} AIC(\mathcal{G}_{t_{b}-1,t_b})
\end{equation}
Where when $t_b = t_{break}(0)$, $t_b-1 = 0$.
In the practical implementation of the algorithm for detecting structural breaks, we start with segments of maximum size of 50 data points, and then move the window forward if no break point is identified. The reason behind this choice is to mitigate potential numerical issues.

From these three models it turns out that:

\begin{center}
\begin{table}[h!]
\centering
	\begin{tabular}{ |c|c|  }
		\hline
		Model & $AIC_{total}$  \\
		\hline
		$\mathcal{H}_{1,nm}$ & 8624.4\\
		$\mathcal{H}_{2,nm}$ & \textbf{1657.11}  \\
		$\mathcal{H}_{3,nm}$ & 1716.39  \\
		\hline
	\end{tabular}
 \caption{In table are shown the performances for the three memory-less models. In bold the model with the best performance.}
 \label{tab_aic_tot_base}
 \end{table}
\end{center}


\newpage
The model with constraints on the degrees emerges as the winner (see Table \ref{tab_aic_tot_base}), indicating that the connection patterns can be more effectively explained by a latent variable associated with each node rather than a single variable for each link. Additionally, it is evident that using a single parameter for the entire model oversimplifies the dynamics, as suggested by the AIC. Therefore, while the formation of individual links may be independent, links that share common nodes exhibit some level of relationship defined by their shared origins.

The next step involves incorporating memory into our model. As mentioned earlier, we can achieve this by considering three different specifications, each coupled with constraints on the time-averaged degree sequence. In this second part, our objective is to evaluate the extent to which the presence of memory is significant, and assess its level of heterogeneity. We adopt the same approach used for the memory-less case, ranking the three "full" models based on the AIC. For each of the three specifications, wwe have that:

\begin{equation}
K_{model} = 
\begin{cases} 
 N + N(N-1)/2& \text{for} \ \mathcal{H}_{1}\\
 N+N & \text{for} \ \mathcal{H}_{2}\\
 N + 1 & \text{for} \ \mathcal{H}_{3}
\end{cases}
\end{equation}

We compare these three models, along with the model that only constrains the degree sequence, using the AIC. It is important to note that the likelihood expression used in this comparison differs from the one mentioned earlier. In this case, we consider not only the models' ability to reproduce the degree distributions, but also their ability to reproduce the persistence of links.

\begin{center}
\begin{table}[h!]
\centering
	\begin{tabular}{ |c|c|  }
		\hline
		Model & $AIC_{total}$  \\
		\hline
		$\mathcal{H}_{2,nm}$ & 1 398 430.36\\
		$\mathcal{H}_{1}$ & 365 824.91  \\
		$\mathcal{H}_{2}$ & \textbf{269 331.03} \\
		$\mathcal{H}_{3}$ & 272 518.09  \\
		\hline
	\end{tabular}
\caption{In table are shown the performances for the memory-less models with local constraints, and the three models with memory. In bold the model with the best performance. }
\label{tab_AIC_memory}
\end{table}
\end{center}

Based on our results presented in table \ref{tab_AIC_memory}, the preferred model is the one with constraints on the persisting degree. This finding suggests that when considering the persistence of links, it is more effective to describe it using variables associated with individual nodes rather than relying on global aspects of the system or link-specific properties.

Since we have information about external events in this particular empirical temporal network, we assess the capability of our best model to capture real changing points. We select the winning model and compare the identified breaks with the known external events.

\begin{figure}[h!]
\includegraphics[width=8cm]{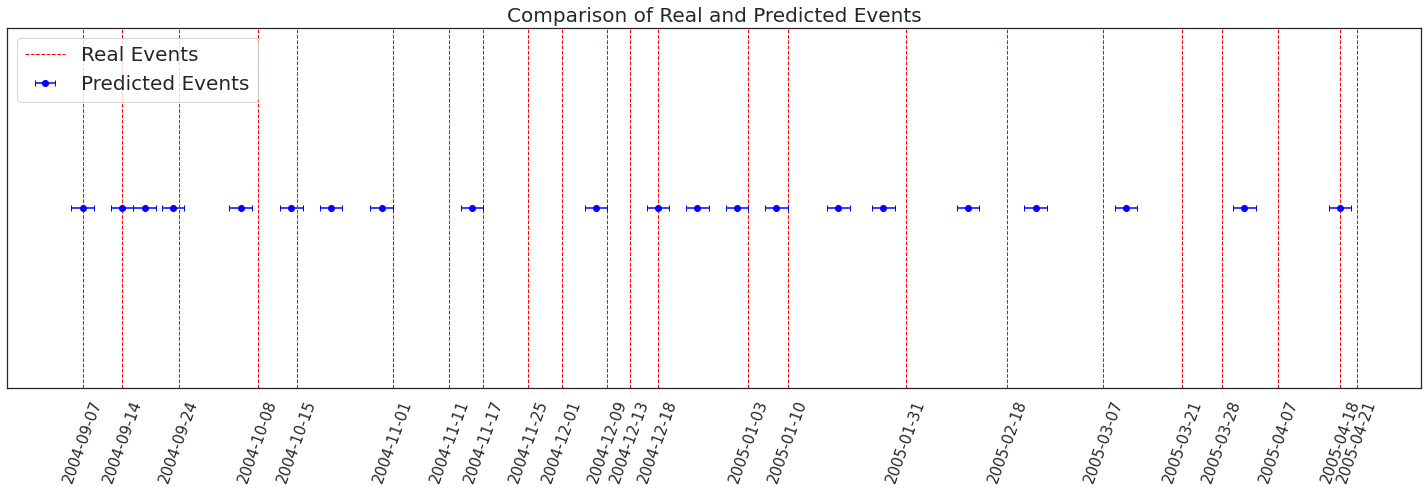}
\centering
\caption{In figure are depicted the points known as real events, in red, and the events predicted in blue.}
\label{structural_breaks}
\end{figure}

In computing the performances, we have to consider that our model works with at least two snapshots, so there is a resolution limitation that must be taken into account: hence we assume that real events are spotted if they last in the interval $t_{break} \pm 2$.
As depicted in figure (\ref{structural_breaks}), we predict a total of 21 events, and within a range of two days, we have been able to spot 11 out of the 23 real events.
Although only some of the events are well captured, it is interesting to observe how many predicted points are not so far from the real ones, highlighting the intrinsic ability of this memory model to capture the effect of exogenous events on the system under analysis.

At this point, we repeat the initial tests on the whole series and compare the persisting degree computed on the shuffled version of the network with the real one for the identified segments using the structural break detection method. We select three of the longest periods found in our data for testing: $T_1 = [2004/10/05 \rightarrow 2004/10/14]$, $T_2=[2004/11/15\rightarrow 2004/12/07]$, and $T_3 = [2005/03/11 \rightarrow 2005/04/01]$.

We observe that as $\tau$ increases, the number of nodes exhibiting a significant positive divergence in the persisting degree compared to the randomization gradually decreases for all three periods, indicating that the system is losing memory of its previous state.

Specifically, the percentage of nodes with a z-score greater than 1.96 (indicating a significant positive difference) decreases from $33\%$ for $\tau=1$ to $0$ for $\tau = 3$ in period $T_1$, from $87\%$ for $\tau=1$ to $3\%$ in period $T_2$, and from $73\%$ to $0$ in $T_3$.
These findings confirm a strong presence of memory among nodes, which we captured using our models, but only with the aid of structural break detection

Now, let's assess the predictive performance of the four models by computing the expected values for $h_i(\tau)$ at different $\tau$ and comparing them with the directly measured values using the Mean Squared Error (MSE) for the three chosen periods.

\begin{equation}
    \text{MSE}(y, \hat{y}) = \frac{1}{n_\text{samples}} \sum_{i=1}^{n_\text{samples}} (y_i - \hat{y}_i)^2.
\label{MSE_equation}
\end{equation}

Where $\hat{y}$ is the predicted value while y, is the true one.

\begin{widetext}
\begin{center}
\begin{table}[h!]
\centering
\begin{tabular}{@{}l|llll|llll|llll|@{}}
\cmidrule(l){2-13}
                          & \multicolumn{4}{l|}{$T_1$}& \multicolumn{4}{l|}{$T_2$}& \multicolumn{4}{l|}{$T_3$}\\ \midrule
\multicolumn{1}{|l|}{$\tau$} & \multicolumn{1}{l|}{$\mathcal{H}_{1}$} & \multicolumn{1}{l|}{$\mathcal{H}_{2}$} & \multicolumn{1}{l|}{$\mathcal{H}_{3}$} & $\mathcal{H}_{2,nm}$ & \multicolumn{1}{l|}{$\mathcal{H}_{1}$} & \multicolumn{1}{l|}{$\mathcal{H}_{2}$} & \multicolumn{1}{l|}{$\mathcal{H}_{3}$} & $\mathcal{H}_{2,nm}$ & \multicolumn{1}{l|}{$\mathcal{H}_{1}$} & \multicolumn{1}{l|}{$\mathcal{H}_{2}$} & \multicolumn{1}{l|}{$\mathcal{H}_{3}$} & $\mathcal{H}_{2,nm}$ \\ \midrule
\multicolumn{1}{|l|}{2}   & \multicolumn{1}{l|}{3.31}& \multicolumn{1}{l|}{\textbf{0.91}}& \multicolumn{1}{l|}{1.13}& 8.74& \multicolumn{1}{l|}{3.07}& \multicolumn{1}{l|}{\textbf{1.33}}& \multicolumn{1}{l|}{2.09}& 17.18& \multicolumn{1}{l|}{0.91}& \multicolumn{1}{l|}{\textbf{0.22}}& \multicolumn{1}{l|}{0.34}& 3.84\\ \midrule
\multicolumn{1}{|l|}{3}& \multicolumn{1}{l|}{2.82}& \multicolumn{1}{l|}{\textbf{1.65}}& \multicolumn{1}{l|}{1.98} & 8.15& \multicolumn{1}{l|}{3.96}& \multicolumn{1}{l|}{\textbf{3.17}} & \multicolumn{1}{l|}{4.17}& 16.16& \multicolumn{1}{l|}{1.04}& \multicolumn{1}{l|}{\textbf{0.48}}& \multicolumn{1}{l|}{0.49}& 3.43\\ \bottomrule
\end{tabular}
\caption{In table are represented the MSE computed between the value of $h_i(\tau)$ predicted with different models vs. the values measured. All tested models were evaluated based on the parameters estimated in the closest segment identified by the structural break algorithm. }
\label{MSE}
\end{table}
\end{center}
\end{widetext}

The results in Table (\ref{MSE}) highlight that, apart from $h_i(\tau=1)$, the models are not required to precisely reproduce the values of $h_i(\tau)$ for $\tau \neq 1$. Notably, the model with constraints on each node consistently outperforms the other models in predicting the actual autocorrelation values for different $\tau$, confirming our previous findings.

It is worth noting that only the memory-less model shows improved performance as $\tau$ increases. This observation suggests a gradual loss of memory as we move further away from the current state of the system. This effect can be better understood by considering the following equations:

\begin{eqnarray}
\langle a_{ij}(t)a_{ij}(t+\tau)\rangle_{m} &=& \langle a_{ij}(t)\rangle_{m}^2 + O(e^{- \frac{\tau}{\tau_{m}^c}}) \label{aut_lm}\\
\langle a_{ij}(t)a_{ij}(t+\tau)\rangle_{nm} &=& \langle a_{ij}(t)\rangle_{nm}^2 
\label{aut_nm}
\end{eqnarray}

Equations (\ref{aut_lm}) and (\ref{aut_nm}) provide a theoretical description of the expected behavior of the autocorrelation function for the models with memory and without memory, respectively. By comparing the results in Table (\ref{MSE}) for the memory-less model, we can infer a non-trivial indication of the Markovian nature of the data.

In equation (\ref{aut_lm}), $\tau_c$ represents the relaxation time, which reflects the tendency to lose memory of the initial condition and is characteristic of each link. It indicates the number of iterations required to reach the stationary distribution.

We know that the relaxation time is related to the second eigenvalue through the expression (\ref{correlation_length_exp}). These quantities can provide further characterization of each node in the system. To achieve this, we compute the second eigenvalue associated with the stochastic matrix of each link in addition to the specifications described by $\mathcal{H}_{2}$

\begin{figure*}
\minipage{0.32\textwidth}
  \includegraphics[width=\linewidth]{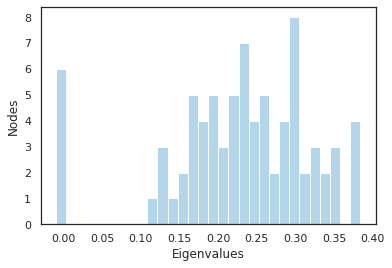}
\endminipage\hfill
\minipage{0.32\textwidth}
  \includegraphics[width=\linewidth]{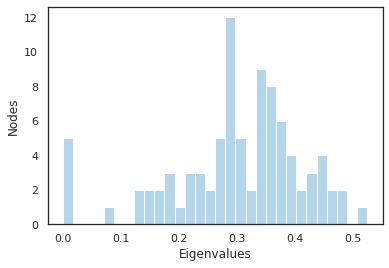}
\endminipage\hfill
\minipage{0.32\textwidth}%
  \includegraphics[width=\linewidth]{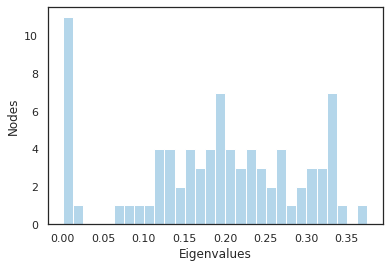}
\endminipage
\label{eigenvalues}
\caption{Here we show the distribution of of average second eigenvalues per node, computed using the model $\mathcal{H}_{2}$. From left to right we consider periods $T_1$, $T_2$ and $T_3$.}
\end{figure*}

In figure (4) is represented the distribution of the average second eigenvalues for each node in three different periods.
For each node, the average is computed by summing all the second eigenvalues defined on the stochastic matrix associated with the links having the node as root, and dividing for the number of nodes.
The distribution in (4), shows a clear difference between nodes and periods remarking a variety that must be taken into account.

Interesting is to observe how the second eigenvalues are differently distributed for different nodes, as reported for three nodes in figure (5).

\begin{figure*}
\minipage{0.32\textwidth}
  \includegraphics[width=\linewidth]{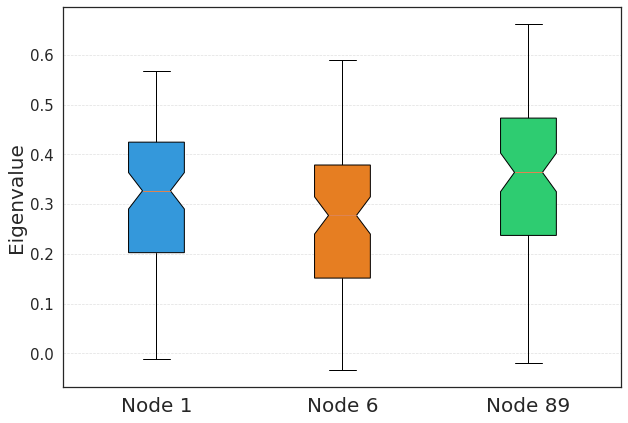}
\endminipage\hfill
\minipage{0.32\textwidth}
  \includegraphics[width=\linewidth]{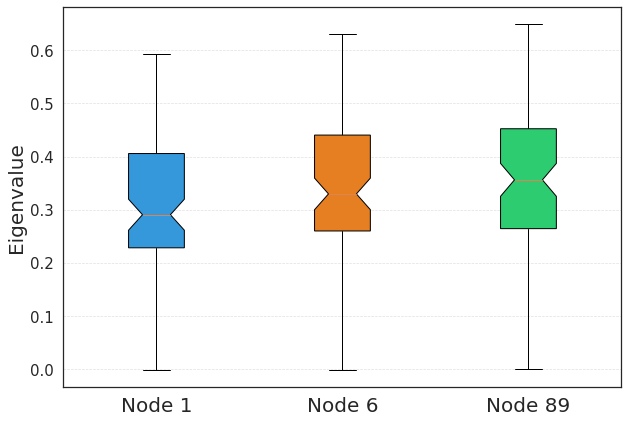}
\endminipage\hfill
\minipage{0.32\textwidth}%
  \includegraphics[width=\linewidth]{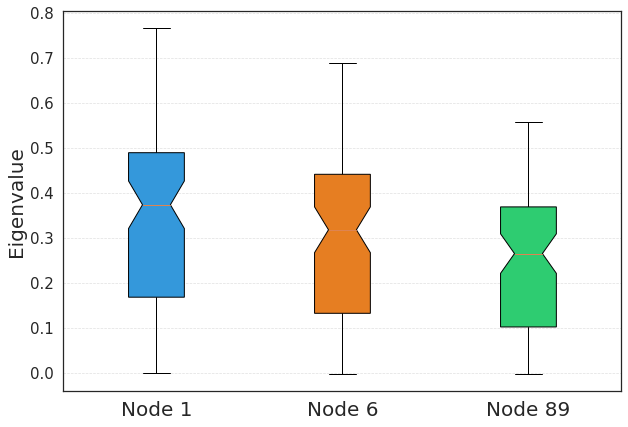}
\endminipage
\label{dist_ave_eigen_and_boxplot}
\caption{The figures display boxplots representing the distributions of second eigenvalues associated with links from three distinct nodes across three separate time periods.}
\end{figure*}

In figure (5), it can be noted how all three elements have a wide and different range of values for the relaxation times. 
It is also evident how different the distributions' shapes are, implying a diversity in the various evolutions that also changes according to the effect of external events, which the model captures well.

\section{Conclusions}
Working with Exponential Random Graph (ERG) models and Temporal Exponential Random Graph (TERG) models often poses challenges in parameter estimation. Estimating the parameters can be problematic due to the analytical summation required to compute the partition function, which may be infeasible or impossible in some cases. These challenges are further compounded in TERG models, which introduce an additional degree of freedom, making tractability even more difficult.

In this article, we addressed these challenges by mapping a specific class of TERGMs, defined using the concept of persisting degree, to a combination of one-dimensional Ising models. This mapping allowed us to derive an analytic solution and calculate the partition function exactly. Consequently, we obtained an exact expression for the probability of having a link between two nodes and the probability of maintaining this link over time.

We demonstrated that having a model that captures memory effects, such as the persisting degree, is crucial for preserving important information. Simplifications like averaging can be useful in certain scenarios but are generally inaccurate. In our framework, we proposed a method to determine when such simplifications are appropriate. Memory can manifest as a global effect, attributable to external causes, or as a local effect specific to each node.

Furthermore, in this study, we investigated the presence of structural breaks in the temporal network under analysis. By employing structural break detection techniques, we identified significant changes in the network's behavior and dynamics. These structural breaks indicate shifts or discontinuities in the underlying processes governing the network's evolution.

We presented a specific application where different memory effects characterize each node, and we showcased how changes in memory can be leveraged for reverse engineering to detect the occurrence of external events. Additionally, we associated each node with a relaxation time, which indicates the number of snapshots required to reach the stationary distribution starting from a random configuration. This concept can be relevant when observing external perturbations in the system and attempting to predict their evolution.

Given the nature of maximum entropy, the null models created within this framework can be applied to various scenarios where an unbiased system for comparison is crucial. This opens up possibilities for discovering hidden patterns in real systems and uncovering new insights.

\section{acknowledgements}
This work is supported by the European Union - NextGenerationEU - National Recovery and Resilience Plan (Piano Nazionale di Ripresa e Resilienza, PNRR), project `SoBigData.it - Strengthening the Italian RI for Social Mining and Big Data Analytics' - Grant IR0000013 (n. 3264, 28/12/2021) (\url{https://pnrr.sobigdata.it/}).
D.G. acknowledges support from the Dutch Econophysics Foundation (Stichting Econophysics, Leiden, the Netherlands) and from the project NetRes - `Network analysis of economic and financial resilience', Italian DM n. 289, 25-03-2021 (PRO3 Scuole), CUP D67G22000130001 (\url{https://netres.imtlucca.it}).

\newpage
\appendix
\section{Supplementary}

\subsection{Finding $\vec{\alpha}$ for the three memory-less models}
Here we show how to derive the functional form of $p_{ij}$ for the three specification of the memory-less model, and how estimate the value of $\vec{\alpha}$ such that the constraints are reproduced.
The most general Hamiltonian between the three is:
\begin{equation}
\mathcal{H}_{1,nm}(\mathcal{G}) = \frac{1}{T}\sum_t \sum_{j<i} \theta_{ij} a_{ij}(t) = \sum_{j<i} \theta_{ij} \overline{a_{ij}}
\label{general_hamiltonian_memory_less}
\end{equation}
We can note that in this case, the constraints are the same used in the work of Newman \cite{Park_Newman} but divided by the number of snapshots of the networks.
This intuition allows us to use the same argument of Newman to obtain the functional form of $p_{ij}$

\begin{equation}
p_{ij} = \langle \overline{a_{ij}} \rangle = \frac{e^{-\theta_{ij}}}{1+e^{-\theta_{ij}}}
\end{equation}
The model defined by (\ref{eq:Hbar_2}) and  (\ref{eq:Hbar_3}), are special cases of this general formulation, where in order, for the specification (\ref{eq:Hbar_2}), $\theta_{ij} = \theta_i + \theta_j$, while for (\ref{eq:Hbar_3}), $\theta_{ij} = \theta/N \ \forall i,j$. The expression of $x_{ij}$ are defined in the main text in eq. (\ref{eq:x_ij}).

To capture the right values of $\vec{\theta}$, such that the constraints are reproduced, we have to solve a series of equations that came from the maximum likelihood principle. 
For the three specifications, the equations are:

\begin{equation}
\langle \overline{a_{ij}} \rangle = \frac{x_{ij}}{1+x_{ij}} = \overline{a_{ij}^*} \qquad \forall i,j \in N
\label{eq:H_nm_1}
\end{equation}
For the first specification (\ref{eq:Hbar_1}). Here we have to solve $\frac{N\cdot(N-1)}{2}$ equations, and $\overline{a_{ij}^*}$ is the average over time of the link between \textit{i} and \textit{j}. 
\begin{equation}
\langle \overline{k_{i}} \rangle = \sum_{i \ne j}\frac{x_{i}x_{j}}{1+x_{i}x_{j}} = \overline{k_{i}^*} \qquad \forall i \in N
\label{eq:H_nm_2}
\end{equation}
For the second expression of the general model (\ref{eq:Hbar_2}). In this case we have to solve $N$ equations, and $\overline{k_{i}^*}$ is the average over time of the degree of node \textit{i}.
\begin{equation}
\frac{x}{1+x} = \frac{1}{N (N-1)} \sum_i \overline{k_{i}^*} = \frac{\overline{k^*}}{(N-1)}
\label{eq:H_nm_3}
\end{equation}
These are the equation for the most general cases among the three, equation (\ref{eq:Ha_3}).
In this last case, the equation to solve is only one, and $\overline{k^*}$ is the average overtime of the average degree.

\subsection{Mapping to the one-dimensional Ising model}
Here we provide a proof of the main results presented in the text. To begin, we explicitly rewrite the Hamiltonian \eqref{eq:Hkh} in the most general case, considering constraints at the level of links ($\mathcal{H}_{1,m}$).

\begin{eqnarray}
\mathcal{H}_1(\mathcal{G})&\equiv& \mathcal{H}_{2,nm}+ \mathcal{H}_{1,m}\nonumber\\
&=&\sum_{i=1}^N\big(\alpha_i \bar{k}_i) + \sum_{j < i} \beta_{ij} \overline{a_{ij}(t) a_{ij}(t+1)} \label{eq:Ha}\\
&=&\frac{1}{T}\sum_{t=1}^T \left[\sum_{i=1}^N\alpha_i k_i(t)+ \sum_{i < j}\beta_{ij}a_{ij}(t) a_{ij}(t+1)\right] \nonumber\\
&=&\frac{1}{T}\sum_{t=1}^T\left[ \sum_{i\ne j}
\alpha_i a_{ij}(t)+ \sum_{j < i}
\beta_{ij} a_{ij}(t)a_{ij}(t+1)\right] \nonumber\\
&=&\frac{1}{T}\sum_{t=1}^T\sum_{j<i}\big[
(\alpha_i+\alpha_j)a_{ij}(t)\nonumber\\&&
\qquad\qquad+
\beta_{ij}a_{ij}(t)a_{ij}(t+1)\big].\nonumber
\end{eqnarray}

The other two cases, as for the memory-less model, are just a particular specification of this, where the two Hamiltonian reads:
\begin{eqnarray}
\mathcal{H}_2(\mathcal{G})&\equiv& \mathcal{H}_{2,nm}+ \mathcal{H}_{2,m}\nonumber\\
&=&\sum_{i=1}^N\big(\alpha_i \bar{k}_i+\beta_i \bar{h}_i(1)\big)\label{eq:Ha_2}
\end{eqnarray}
Using $\mathcal{H}_{2,m}$, and
\begin{eqnarray}
\mathcal{H}_3(\mathcal{G}) &\equiv&\mathcal{H}_{2,nm}+ \mathcal{H}_{3,m}\nonumber\\
&=&\left(\sum_ {i=1}^N\alpha_i \bar{k}_i\right) +\beta \bar{h} \label{eq:Ha_3}
\end{eqnarray}
for $\mathcal{H}_{3,m}$.\\
From now on, we will refer to $\mathcal{H}_1(\mathcal{G})$ as $\mathcal{H}(\mathcal{G})$.

We now focus on eq.(\ref{eq:Ha}). Let us introduce the `spin' variables $\sigma_{ij}(t)=\pm 1$ defined through
\begin{equation}
a_{ij}(t)\equiv\frac{\sigma_{ij}(t)+1}{2}.
\label{eq:a}
\end{equation}
In eq.\eqref{eq:Ha}, each graph $\mathbf{G}(t)$ in the trajectory $\mathcal{G}$ has been parametrized by the adjacency matrix $\mathbf{A}(t)$ with entries $a_{ij}(t)$. Equivalently, we can parametrize $\mathbf{G}(t)$ by the `spin matrix' $\mathbf{\Sigma}(t)$ with entries $\sigma_{ij}(t)$.
If we insert eq.\eqref{eq:a} into eq.\eqref{eq:Ha}, after some algebra we arrive at
\begin{eqnarray}
\mathcal{H}(\mathcal{G})=\mathcal{H}_0-\sum_{t=1}^T\sum_{j<i}\big[
B_{ij}\sigma_{ij}(t)+J_{ij}\sigma_{ij}(t)\sigma_{ij}(t+1)\big]\nonumber
\end{eqnarray}
where we have defined
\begin{eqnarray}
B_{ij}&\equiv& -\frac{1}{2T}(\alpha_i+\alpha_j+\beta_{ij})\\
J_{ij}&\equiv& -\frac{1}{4T}(\beta_{ij})\\
\mathcal{H}_0&\equiv&\sum_{j<i}
\left(\frac{\alpha_i+\alpha_j}{2}
+\frac{\beta_{ij}}{4}\right).
\end{eqnarray}

Note that $\mathcal{H}_0$ is a constant term that does not depend on the particular graph trajectory $\mathcal{G}$. It has therefore no effect on any observable of the model, and we can drop it.
The Hamiltonian then reads
\begin{equation}
\mathcal{H}(\mathcal{G})=\sum_{j<i}H(\mathcal{G}_{ij})
\label{eq:Hsum}
\end{equation}
where, if we still parametrize $\mathbf{G}(t)$ using the matrix $\mathbf{\Sigma}(t)$, $\mathcal{G}_{ij}=\{\sigma_{ij}(1),\dots,\sigma_{ij}(T)\}$ is the restriction of $\mathcal{G}$ to the pair of nodes $i,j$, i.e. the temporal sequence of (connection/disconnection) events for such pair, and
\begin{equation}
H(\mathcal{G}_{ij})\equiv -\sum_{t=1}^T \Big[ B_{ij}\sigma_{ij}(t)+J_{ij}\sigma_{ij}(t)\sigma_{ij}(t+1)\Big].
\label{eq:Hij}
\end{equation}

A crucial observation is that eq. \eqref{eq:Hij} coincides with the Hamiltonian of a one-dimensional Ising model with first-neighbour interactions, where each `spin' $\sigma_{ij}(t)$ sits on the lattice site $t$ and interacts with the spin on the site $t+1$. The parameters $B_{ij}$ and $J_{ij}$ play the role of the external field and coupling constant respectively.
As well known, this model can be solved analytically \cite{baxter2007exactly}. Before showing the solution, we note that eq. \eqref{eq:Hsum} implies that if
\begin{equation}
Z(B_{ij},J_{ij})\equiv\sum_{\mathcal{G}_{ij}} e^{-H(\mathcal{G}_{ij})}
\label{eq:Zij}
\end{equation}
is the partition function for the dyad-specific Hamiltonian \eqref{eq:Hij}, then the partition function for the graph-wide Hamiltonian \eqref{eq:Ha} is
\begin{equation}	
\mathcal{Z}(\mathbf{B},\mathbf{J})\equiv \sum_{\mathcal{G}} e^{-\mathcal{H}(\mathcal{G})}=\prod_{j<i}Z(B_{ij},J_{ij}).
\label{eq:Z}
\end{equation}
Similarly, if 
\begin{equation}
P(\mathcal{G}_{ij}|B_{ij},J_{ij})=\frac{e^{-H(\mathcal{G}_{ij})}}{Z(B_{ij},J_{ij})}
\end{equation}
is the probability of the dyadic trajectory $\mathcal{G}_{ij}$ in the model defined by eq. \eqref{eq:Hij}, then the probability $\mathcal{P}(\mathcal{G}|\mathbf{B},\mathbf{J})$ of an entire graph trajectory $\mathcal{G}$ in the model specified by eq.\eqref{eq:Ha} is
\begin{equation}
\mathcal{P}(\mathcal{G}|\mathbf{B},\mathbf{J})=\frac{e^{-\mathcal{H}(\mathcal{G})}}{\mathcal{Z}(\mathbf{B},\mathbf{J})}=\prod_{j<i}P(\mathcal{G}_{ij}|B_{ij},J_{ij})
\label{eq:P}
\end{equation}

\subsection{Partition function}
We now show the solution of the dyadic model defined by eq. \eqref{eq:Hij}, by adapting it from the known solution of the one-dimensional Ising model \cite{baxter2007exactly}. 
The periodicity condition \eqref{eq:periodic} ensures that all sites (time steps) are statistically equivalent, i.e.
\begin{equation} 
 \langle \sigma_{ij}(1)\rangle=\langle \sigma_{ij}(2)\rangle=\dots=\langle \sigma_{ij}(T)\rangle
\end{equation}
The system is therefore translationally (here, temporally) invariant. 
The partition function \eqref{eq:Zij} is  
\begin{equation} 
Z(B_{ij},J_{ij}) =  \sum_{\mathcal{G}_{ij}} \exp\Big[B_{ij}  \sigma_{ij}(t) + J_{ij} \sigma_{ij}(t)\sigma_{ij}(t+1) \Big]
\end{equation}
and can be rewritten as a product of terms involving only two successive time steps:
\begin{equation} 
Z(B_{ij},J_{ij}) = \prod_{t=1}^T \sum_{\sigma_{ij}} V_{ij}\big(\sigma_{ij}(t),\sigma_{ij}(t+1)\big),
\label{eq:partition}
\end{equation} 
where the function $V_{ij}(x,y)$ is defined as
\begin{equation} 
V_{ij}(x,y)\equiv\exp\left(\frac{x+y}{2}B_{ij}+xyJ_{ij}\right).
\end{equation}

Since both $x$ and $y$ can take only the values $\pm 1$, we can regard the four quantities $V_{ij}^{\pm \pm}\equiv V_{ij}(\pm1,\pm1)$ as the entries of a $2\times2$ matrix $\mathbf{V}_{ij}$ called the \emph{transfer matrix} \cite{baxter2007exactly}:
\begin{equation} 
\mathbf{V}_{ij}\equiv \left( \begin{array}{cc} V_{ij}^{++} &  V_{ij}^{+-} \\ V_{ij}^{-+}&  V_{ij}^{--} \end{array}\right)
=\left( \begin{array}{cc} e^{J_{ij}+B_{ij}}&  e^{-J_{ij}} \\  e^{-J_{ij}}&  e^{J_{ij}-B_{ij}} \end{array}\right).
\end{equation} 
This allows us to rewrite eq.\eqref{eq:partition} as 
\begin{equation} 
Z(B_{ij},J_{ij}) = \textrm{Tr} \big(\mathbf{V}_{ij}^T\big)
\label{eq:trace}
\end{equation}
(where $\mathbf{V}_{ij}^T$ denotes the $T$-th matrix power of $\mathbf{V}_{ij}$, and \emph{not} the transpose of the latter). 

Now let $\vec{v}_{ij}^{~\pm}$ denote the two eigenvectors of $\mathbf{V}_{ij}$, and $\lambda_{ij}^{\pm}$ the corresponding eigenvalues (with $\lambda_{ij}^{+}\ge\lambda_{ij}^{-}$), so that 
\begin{equation}
\mathbf{V}_{ij}\vec{v}_{ij}^{~\pm}=\lambda_{ij}^{\pm}\vec{v}_{ij}^{~\pm}.
\end{equation}
The $2\times 2$ matrix $\mathbf{Q}_{ij}\equiv (\vec{v}_{ij}^{~+},\vec{v}_{ij}^{~-})$, having column vectors $\vec{v}_{ij}^{~+}$ and $\vec{v}_{ij}^{~-}$, diagonalizes $\mathbf{V}_{ij}$, i.e.  
\begin{equation} 
\mathbf{V}_{ij}=\mathbf{Q}_{ij}  \left( \begin{array}{cc} \lambda_{ij}^{+} &  0 \\ 0&  \lambda_{ij}^{-} \end{array}\right) \mathbf{Q}_{ij}^{-1}.
\end{equation}
A direct calculation of the eigenvalues and eigenvectors yields
\begin{equation}
\lambda_{ij}^{\pm} = e^{J_{ij}} \cosh B_{ij}\pm\sqrt{e^{2J_{ij}}\sinh^2 B_{ij}+e^{-2J_{ij}}}
\end{equation}


It then follows, using the cyclic properties of the trace, that eq.\eqref{eq:trace} simply reduces to 
\begin{equation} 
Z(B_{ij},J_{ij}) = \textrm{Tr}\left( \begin{array}{cc} \lambda_{ij}^{+} &  0 \\ 0&  \lambda_{ij}^{-} \end{array}\right)^{T} = \big(\lambda_{ij}^{+}\big)^{T} +\big(\lambda_{ij}^{-}\big)^{T}
\end{equation}
and the full partition function, eq. \eqref{eq:Z}, is
\begin{equation}
\mathcal{Z}(\mathbf{B},\mathbf{J})=\prod_{j<i}\Big[\big(\lambda_{ij}^{+}\big)^{T} +\big(\lambda_{ij}^{-}\big)^{T}\Big].
\end{equation}

Similarly, the probability of the dyadic trajectory $\mathcal{G}_{ij}$ is 
\begin{equation}
P(\mathcal{G}_{ij}|B_{ij},J_{ij})=\frac{\prod_{t=1}^T V_{ij}\big(\sigma_{ij}(t),\sigma_{ij}(t+1)\big)}{\big(\lambda_{ij}^{+}\big)^{T} +\big(\lambda_{ij}^{-}\big)^{T}}
\end{equation}
and that of the entire graph trajectory $\mathcal{G}$ is 
\begin{equation}
\mathcal{P}(\mathcal{G}|\mathbf{B},\mathbf{J})=\prod_{j<i}\frac{\prod_{t=1}^T V_{ij}\big(\sigma_{ij}(t),\sigma_{ij}(t+1)\big)}{\big(\lambda_{ij}^{+}\big)^{T} +\big(\lambda_{ij}^{-}\big)^{T}}
\label{eq:PP}
\end{equation}

\subsection{Expected values}

We start calculating the expected value of $\sigma_{ij}(t)$ in two different ways.\\
We do this because we will use some of the passages in both the calculations to compute the autocorrelation function.

In the first approach we use the free energy per spin, defined by the following equation:
\begin{equation}
	F_{ij} = -\frac{1}{T} ln Z(B_{ij},J_{ij}) 
\end{equation}
And differentiating $F_{ij}$ with respect to $B_{ij}$, we obtain $\langle \sigma_{ij}(t)\rangle$.

\begin{eqnarray}
&\frac{\partial ln Z(B_{ij},J_{ij}) }{\partial B_{ij}} = \frac{1}{(\lambda_{ij}^{+})^T + (\lambda_{ij}^{-})^T} \bigg[  T (\lambda_{ij}^+)^{T-1}\frac{\partial \lambda_{ij}^+}{\partial B_{ij} } + &  \nonumber \\
&+ T (\lambda_{ij}^-)^{T-1}\frac{\partial \lambda_-}{\partial B_{ij} }   \bigg]& 
\end{eqnarray}

Where $\frac{\partial \lambda_{ij}^{\pm}} {\partial B_{ij}}$ has the following expression,

\begin{equation}
\frac{\partial \lambda_{ij}^{\pm}} {\partial B_{ij}} = e^{J_{ij}} \left[  sinhB_{ij} \pm \frac{sinhB_{ij} coshB_{ij}}{ \sqrt{sinh(B_{ij})^{2} + e^{-4J_{ij}} } }      \right]
\end{equation}

Considering that
\begin{equation}
	 \langle \sigma_{ij}(t) \rangle =  - \frac{\partial F_{ij}}{\partial B_{ij}}
\end{equation}

We can finally obtain the expected value of $\sigma_{ij}(t)$

\begin{eqnarray}
 &\langle \sigma_{ij}(t) \rangle =& 	\nonumber \\ 
 &\frac{e^{J_{ij}} } {(\lambda_{ij}^+)^T + (\lambda_{ij}^-)^T}  \Bigg[  (\lambda_{ij}^+)^{T-1}   \left(  sinhB_{ij} + \frac{sinhB_{ij} \ coshB_{ij} }{ \sqrt{sinh(B_{ij})^2 + e^{-4J_{ij}} } }   \right)  +   & \nonumber \\
   & +(\lambda_{ij}^-)^{T-1} \left(   sinhB_{ij} - \frac{sinhB_{ij} \ coshB_{ij}}{ \sqrt{sinh(B_{ij})^2 + e^{-4J_{ij}} } }    \right)   \Bigg]  &
\end{eqnarray}

Given this general result, we  can move to the thermodynamic limit, obtaining a simplified expression:
\begin{eqnarray}
	\langle \sigma_{ij}(t) \rangle_{tl} = \frac{e^{J_{ij}}}{\lambda_{ij}^+}  \left[    sinhB_{ij}+  \frac{sinhB_{ij} \ coshB_{ij}}{ \sqrt{sinh(B_{ij})^{2} + e^{-4J_{ij}} } } \right] =  \nonumber & \\ \frac{sinhB_{ij}}{\sqrt{sinh(B_{ij})^{2} + e^{-4J_{ij}} }} \hspace{10mm}
	\label{sigmafreeEn}
\end{eqnarray}  

As anticipated, the result obtained in eq. \ref{sigmafreeEn} can be procured using another technique: The transfer matrix method \cite{baxter2007exactly}. \\
By introducing S, which is a diagonal matrix such that on its diagonal there are all the possible spin values:

\begin{equation}
\mathbf{S}\equiv\left( \begin{array}{cc} S^{++} &  S^{+-} \\ S^{-+}&  S^{--} \end{array}\right) 
=\left( \begin{array}{rr} +1 & 0 \\
0& -1 \end{array} \right)
\end{equation}
having elements $S^{\pm\pm}\equiv S(\pm1,\pm1)$, with
\begin{equation}
S(x,y)\equiv x \delta (x,y).
\end{equation}

We can rewrite $\langle \sigma_{ij}(t)\rangle$:
\begin{eqnarray} 
\langle \sigma_{ij}(t)\rangle&\equiv&
\sum_{\mathcal{G}_{ij}}\sigma_{ij}(t)P(\mathcal{G}_{ij}|B_{ij},J_{ij})\nonumber\\
&=&
 \frac{\textrm{Tr}\big(\mathbf{S}\mathbf{V}_{ij}^T\big)}{\big(\lambda_{ij}^{+}\big)^{T} +\big(\lambda_{ij}^{-}\big)^{T}},
\label{eq:x}
\end{eqnarray}

And using the eigenvector of the transfer matrix, we have:
\begin{eqnarray}
	\langle \sigma_{ij}(t)\rangle =  &\frac{1}{Z(B_{ij},J_{ij})} \bigg[ (\lambda_{ij}^+)^T <v_{ij}^+|S|v_{ij}^+ > +&\\
	&+ (\lambda_{ij}^-)^T<v_{ij}^-|S|v_{ij}^->\bigg] &\\
	\label{sigmanolim}
\end{eqnarray}
That, using also the results \ref{sigmafreeEn}, in the thermodynamic limit is:
\begin{eqnarray}
		\langle \sigma_{ij}(t) \rangle_{tl} = \langle v_{ij}^+|S|v_{ij}^+ \rangle = &\\ 
		=\frac{e^{J_{ij}}}{\lambda_{ij}^+}  \left[    sinhB_{ij} +  \frac{sinhB_{ij} \ coshB_{ij}}{ \sqrt{sinh(B_{ij})^{2} + e^{-4J_{ij}} } } \right]
		\label{sigmalim}
\end{eqnarray}

Hence, putting together (\ref{sigmalim}), (\ref{sigmanolim}) and (\ref{sigmafreeEn}), we obtain an expression for the quantity $\langle v_{ij}^-|S|v_{ij}^- \rangle$
\begin{multline}
		\langle v_{ij}^-|S|v_{ij}^- \rangle = \\ 
		\frac{e^{J_{ij}}}{\lambda_{ij}^-}  \left[    sinhB_{ij} -  \frac{sinhB_{ij} \ coshB_{ij}}{ \sqrt{sinh(B_{ij})^{2} + e^{-4J_{ij}} } } \right] =\\
		 - \langle v_{ij}^+|S|v_{ij}^+ \rangle
		 \label{v^-_ij}
\end{multline}

And after some manipulations, we have that for a finite number of snapshots, the expected value of a spin is:
\begin{equation}
	\langle \sigma_{ij}(t) \rangle = \langle v_{ij}^+|S|v_{ij}^+ \rangle \left( \frac{(\lambda_{ij}^+)^T - (\lambda_{ij}^-)^T}{(\lambda_{ij}^+)^T + (\lambda_{ij}^-)^T} \right)
	\label{nonTLsigma}
\end{equation}

Now we have all the elements to compute the expected value of $\sigma_{ij}(t)\sigma_{ij}(t+\tau)$, for $ 0<  \tau < T$.\\
We note that similar to the expected value of the spin function it can be expressed using the transfer matrix:

\begin{eqnarray}
		&\langle \sigma_{ij}(t) \sigma_{ij}(t+\tau) \rangle = \frac{1}{Z(B_{ij},J_{ij})}  \sum_{ \{\sigma_i=\pm 1 \}}  \sigma_t \langle \sigma_t|V|\sigma_{t+1} \rangle \dots & \nonumber \\
		&\langle  \sigma_{t+\tau - 1} |V|\sigma_{t+\tau} \rangle \sigma_{t+\tau} \langle \sigma_{t+\tau} |V|\sigma_{t+\tau + 1} \rangle \dots = & \nonumber \\
		&\frac{1}{Z(B_{ij},J_{ij})} \sum_{\sigma_t,\sigma_{t+\tau}} \sigma_t \langle \sigma_t|V^{\tau}|\sigma_{t+\tau} \rangle \sigma_{t+\tau} \langle \sigma_{t+\tau}|V^{T-\tau}|\sigma_1 \rangle& \nonumber\\
\end{eqnarray}
 
 It is also possible to write matrix V in its eigenvector space $|v_i \rangle$:
 \begin{equation}
 V = \sum_{l \in \{+,- \} } |v_l \rangle \lambda_l \langle v_l|
 \end{equation}
 
 Where $\lambda_i$ are the correspondent eigenvalues.
 
 In this way, given that the eigenvectors are orthonormal we have:
 \begin{equation}
 V^{n} = \sum_{l \in \{+,- \}} |v_l \rangle \lambda_l^{n} \langle v_l|
 \end{equation}
 
 Hence:
 \begin{eqnarray}
	\langle \sigma_{ij}(t) \sigma_{ij}(t+\tau) \rangle  = \frac{1}{Z(B_{ij},J_{ij})}  \sum_{ \sigma_t, \sigma_{t+\tau} } \sum_{i,j \in \{+,- \}} \bigg[\sigma_t \langle\sigma_t |v_i \rangle  \nonumber\\ \lambda_i^{t+\tau}\langle v_i|\sigma_{t+\tau}  \rangle   \sigma_{t+\tau} \langle \sigma_{t+\tau}|v_j  \rangle    \lambda_j^{T-\tau}  \langle v_j|\sigma_t \rangle \bigg] \hspace{9mm}&
	\label{eq_no_s_i}
 \end{eqnarray}
 
 We can then rewrite eq. (\ref{eq_no_s_i}) by using the matrices $S_i$, using its the eigenvectors base:
 \begin{equation}
	 S_i = \sum_{\sigma_i} |\sigma_i\rangle \sigma_i \langle \sigma_i| 
 \end{equation}
 In this way moving $\langle v_j|\sigma_t\rangle $ (is just a number) and summing over $\sigma_{ij}(t)$ and $\sigma_{ij}(t+\tau)$, we obtain:
 
 \begin{equation}
 \langle \sigma_{ij}(t) \sigma_{ij}(t+\tau) \rangle  = \frac{ \sum_{i,j} \langle  v_j|S_t|v_i \rangle \lambda_i^{\tau} \langle v_i|S_{t+\tau}|v_j \rangle \lambda_j^{T-\tau}     }{\sum_{i} \lambda_i^{T}}
 \label{correlation}
 \end{equation}
 
 Where $S_i = S \ \forall i$.
 
 Hence, after writing explicitly the summation, eq.(\ref{correlation}) becomes:
 \begin{eqnarray}
		&\langle \sigma_{ij}(t) \sigma_{ij}(t+\tau) \rangle  = \frac{1}{\lambda_+^{T} + \lambda_-^{T}} \bigg[  \lambda_+^{T} \langle v^{+}|S|v^{+} \rangle^2+ \lambda_-^{T} \langle v^{-}|S|v^{-} \rangle^{2} + \nonumber \\
		 & + ((\lambda_{ij}^-)^{\tau}  (\lambda_{ij}^+)^{T-\tau}   +(\lambda_{ij}^+)^{\tau} (\lambda_{ij}^-)^{T-\tau}   )\langle v^{+}|S|v^{+} \rangle (1-  \langle v^{+}|S|v^{+} \rangle   )  \bigg] & \nonumber \\
 \end{eqnarray}
 
And using the results of eq.(\ref{v^-_ij}) and (\ref{sigmalim}), after some algebraical passages we get:
\begin{widetext}
\begin{equation}
\langle \sigma_{ij}(t) \sigma_{ij}(t+\tau) \rangle = \frac{1}{Z} Tr\left[    SV^{\tau} S V^{N-\tau}  \right] = 
\langle \sigma_{ij}(t) 	\rangle_{tl}^2 +\langle  \sigma_{ij}(t) \rangle_{tl}(1-\langle \sigma_{ij}(t) \rangle_{tl})\left(  \frac{(\lambda_{ij}^-)^\tau (\lambda_{ij}^+)^{T-\tau}  + (\lambda_{ij}^+)^\tau (\lambda_{ij}^-)^{T-\tau}    }{(\lambda_{ij}^+)^T + (\lambda_{ij}^-)^T} \right)
\label{correlation_full}
\end{equation}
 \end{widetext}

That in the limit $T\to\infty$ (corresponding to long graph trajectories) with $\tau$ fixed and finite (which necessarily implies $\tau\ll T$)  becomes
 
\begin{eqnarray}
	&\langle \sigma_{ij}(t) \sigma_{ij}(t+\tau) \rangle = \langle \sigma(t) \rangle_{tl}^2 + \langle \sigma(t) \rangle_{tl}(1-\langle \sigma(t)\rangle_{tl})\left(  \frac{\lambda_{ij}^+}{\lambda_{ij}^-} \right)^{\tau}& \nonumber \\ 
\end{eqnarray}

Now, it should be noted that the model manifestly shows a `translational', which in our case means \emph{temporal}, invariance, as both $\langle \sigma_{ij}(t)\rangle$ and $\langle \sigma_{ij}(t)\sigma_{ij}(t+\tau)\rangle$ are independent of $t$.
This implies that the expectation value of the time-averaged quantities concides with that of the quantities themselves:
\begin{eqnarray}
\langle \overline{\sigma_{ij}(t)}\rangle&=& \langle {\sigma_{ij}(t)}\rangle \\
\langle \overline{\sigma_{ij}(t) \sigma_{ij}(t+\tau) }\rangle&=& \langle {\sigma_{ij}(t) \sigma_{ij}(t+\tau) }\rangle
\end{eqnarray}
A similar result holds if we come back to the variables $a_{ij}(t)$ through eq. \eqref{eq:a}, so that
\begin{eqnarray}
p_{ij}&\equiv&\langle \overline{a_{ij}(t)}\rangle = \langle {a_{ij}(t)}\rangle\label{eq:pexp}\\
q_{ij}(\tau)&\equiv& \langle \overline{a_{ij}(t)a_{ij}(t+\tau)}\rangle = \langle {a_{ij}(t)a_{ij}(t+\tau)}\rangle\qquad
\end{eqnarray}

Therefore, using eqs. \eqref{eq:a}, and \eqref{nonTLsigma} for $p_{ij}$ we get
\begin{eqnarray}
&p_{ij}&= \left(\frac{e^{2J_{ij}}\sinh{B_{ij}}}{2\sqrt{1+e^{4J_{ij}}\sinh^2{B_{ij}}}}\right)\left( \frac{(\lambda_{ij}^+)^T - (\lambda_{ij}^-)^T}{(\lambda_{ij}^+)^T + (\lambda_{ij}^-)^T} \right) +\frac{1}{2} \nonumber\\
&=&\left( \frac{x_i x_j y_{ij}-1}{2\sqrt{4x_i x_j+(x_i x_j y_{ij} -1)^2}} \right)\left( \frac{(\lambda_{ij}^+)^T - (\lambda_{ij}^-)^T}{(\lambda_{ij}^+)^T + (\lambda_{ij}^-)^T} \right) +\frac{1}{2}, \nonumber\\ 
\label{eq:px}
\end{eqnarray}
where we have introduced the parameters
\begin{equation} 
x_i\equiv e^{-\alpha_i},\quad y_{ij}\equiv e^{-\beta_{ij}}.
\end{equation}

While for $q_{ij}(\tau)$, a similar derivation leads to 
\begin{eqnarray}
&q_{ij}(\tau)=& \nonumber \\ &\tilde{p}_{ij}^2+\tilde{p}_{ij} (1-\tilde{p}_{ij}) \left(  \frac{(\lambda_{ij}^-)^\tau (\lambda_{ij}^+)^{T-\tau}  + (\lambda_{ij}^+)^\tau (\lambda_{ij}^-)^{T-\tau}    }{(\lambda_{ij}^+)^T + (\lambda_{ij}^-)^T} \right)\qquad& \nonumber \\ 
\label{eq:qx}
\end{eqnarray}

Where $\tilde{p}_{ij}$ corresponds to the value of $p_{ij}$ computed in the long limit trajectory.

\begin{eqnarray}
\tilde{p}_{ij}= \left(\frac{e^{2J_{ij}}\sinh{B_{ij}}}{2\sqrt{1+e^{4J_{ij}}\sinh^2{B_{ij}}}}\right) +\frac{1}{2} 
\label{eq:px_lim}
\end{eqnarray}

And $q_{ij}$ in the thermodynamic limit assumes the following expression:
\begin{eqnarray}
\tilde{q}_{ij}(\tau)= \tilde{p}_{ij}^2+\tilde{p}_{ij} (1-\tilde{p}_{ij}) \left(  \frac{\lambda_{ij}^+}{\lambda_{ij}^-} \right)^{\tau} 
\label{eq:qx_lim}
\end{eqnarray}

\subsection{Maximum likelihood estimation}
The above expressions allow us to calculate all the relevant expected properties of the time series generated by the model, once the parameters $\mathbf{B}$ and $\mathbf{J}$ 
are set to the values $\mathbf{B}^*$ and $\mathbf{J}^*$ maximizing the likelihood $\mathcal{P}(\mathcal{G}^*|\mathbf{B},\mathbf{J})$ of the observed graph trajectory $\mathcal{G}^*$, where $\mathcal{P}(\mathcal{G}|\mathbf{B},\mathbf{J})$ is given by eq. \eqref{eq:PP}.
Equivalently, in terms of the parameters $\vec{x}$ and $\vec{y}$ we look for the values $\vec{x}^*$ and $\vec{y}^*$ that maximize the same quantity.

As we mentioned in the main text, since $\mathcal{P}(\mathcal{G}|\mathbf{B},\mathbf{J})$ has the form \eqref{eq:P} where $\mathcal{H}(\mathcal{G})$ is given by eq. \eqref{eq:Ha}, a general theorem \cite{likelihood_Garlaschelli_loffredo} ensures that the  values $\vec{x}^*$ and $\vec{y}^*$ that maximize the likelihood are those that solve the $N + \frac{(N\cdot(N-1))}{2}$ equations
\begin{eqnarray}
\langle \bar{k}_i\rangle&=&\sum_{j\ne i}p_{ij}=\bar{k}^*_i\quad\forall i
\label{eq:khbarconstraint_1}\\
\langle \overline{a_{ij}(t)a_{ij}(t+1)}\rangle&=& q_{ij}(1)=\bar{h}^*_{ij}\quad\forall i,j
\label{eq:khbarconstraint_2}
\end{eqnarray}
where $p_{ij}$ and $q_{ij}(1)$, which are both functions of $\vec{x}$ and $\vec{y}$, are given by eqs. \eqref{eq:px} and \eqref{eq:qx} respectively.

For the two particular cases defined by $\mathcal{H}_{3}$ and $\mathcal{H}_{3}$, we have that in place of the $\frac{(N\cdot(N-1))}{2}$ eqs. (\ref{eq:khbarconstraint_2}) we need to solve, either the $N$ equations, given by:
\begin{equation}
\langle \bar{h}_i\rangle = \sum_{j\ne i}q_{ij}(1)=\bar{h}^*_i\quad\forall i
\end{equation}

Or the single equation given by:

\begin{equation}
 \frac{1}{N} \sum_i^N \langle \bar{h}_i\rangle = \sum_{i \ne j}q_{ij}(1)=\frac{1}{N} \sum_i^N \bar{h}^*_i
\end{equation}

For this last case, it is worth to note that even if we constraint $\beta$ to be equal for all nodes, $q_{ij}$ is still different, because of its dependence on the value of $\vec{\alpha}$.

In general, the expressions for $p_{ij}$ and $q_{ij}$ are the ones described in eqs. (\ref{eq:px}) and (\ref{eq:qx}), although if the long trajectory limits is satisfied, the thermodynamic limit versions can be implemented.

\subsection{From Memory to Memory-Less}

Here we show how, starting form the full model and turning off the memory of the system we obtain the same expression of $p_{ij}$ in the memory-less case.
We note that the case of no memory can be translated by imposing $\beta_i = 0 \ \forall i$, that is $J_{ij}= 0 \ \forall i, j $. \\ 
This implies that:
\begin{equation}
\lambda_{ij}^+=2 coshB_{ij};
\end{equation}
\begin{equation}
\lambda_{ij}^- = 0
\end{equation}
\begin{equation}
\sqrt{sin(B_{ij})^{2} + e^{-4J_{ij}}} = coshB_{ij}
\end{equation}

Using these equations in the expression of $\langle \sigma_{i,j}(t) \rangle $ (eq. (\ref{nonTLsigma})), we have:
\begin{equation}
\langle \sigma_{ij}(t) \rangle = \frac{sinhB_{ij}}{coshB_{ij}} = \frac{e^{2B_{ij}} - 1}{e^{2B_{ij}} + 1}
\end{equation}

Considering that $B_{ij}=\frac{1}{2} [ln (x_i) + ln (x_j)]$, we obtain

\begin{equation}
\langle \sigma_{ij}(t) \rangle=\left(   \frac{x_i x_j -1}{x_i x_j +1}  \right)
\end{equation}

The quantity of interest is  $ \langle a_{ij} \rangle $ and recalling that $  \langle a_{ij} \rangle = \frac{\langle \sigma_{t}\rangle   + 1}{2}$, we get:
\begin{equation}
\langle a_{ij} \rangle =  \frac{x_i x_j }{x_i x_j +1}
\end{equation}  

That corresponds exactly to the expected value of $a_{ij}$ in the memory-less case.

\subsection{Deriving The Stochastic Matrices}

The model is built such that each link is associated with a stochastic matrix. In this section, we describe how this matrix is defined.\\
We start using the quantities defined in the model connecting them in order to introduce the coupled probabilities of events for each link:

1) The probability that two nodes have a connection at time t and $t+\tau$ (with $\tau=1$):
\begin{eqnarray}
P[a_{ij}(t+\tau) = 1 \ \& \ a_{ij}(t)=1   ] = & \nonumber \\
\langle a_{ij}(t) a_{ij}(t+1) \rangle =& q_{ij} &   
\end{eqnarray}

2) The probability that two nodes have a connection at time t and not $t+\tau$:
\begin{eqnarray}
P[a_{ij}(t+\tau) = 0  \ \& \ a_{ij}(t)=1   ] = &\nonumber \\
   \langle a_{ij}(t)(1- a_{ij}(t+1)) \rangle =& p_{ij} - q_{ij}& 
\end{eqnarray}

3) The probability that two nodes don't have a connection at time t having a connection a time $t+\tau$:
\begin{eqnarray}
P[a_{ij}(t+\tau) = 1 \ \& \ a_{ij}(t)=0   ] =& \nonumber \\
 \langle (1-a_{ij}(t) ) a_{ij}(t+1) \rangle =& p_{ij} - q_{ij}&
\end{eqnarray}

4) The probability that two nodes don't have a connection at time t and $t+\tau$:
\begin{eqnarray}
P[a_{ij}(t+\tau) = 0 \ \& \ a_{ij}(t)=0   ] = &  \nonumber\\
\langle (1-a_{ij}(t) ) (1-a_{ij}(t+1)) \rangle = & 1 - 2 p_{ij} +q_{ij} &
\end{eqnarray}

Defined these elements, we can write the transition matrix $P_{ij}$, equal in the form for each couple of nodes:

\begin{equation}
P_{ij} = 
\begin{bmatrix}
\frac{q_{ij}}{p_{ij}} & \frac{p_{ij}-q_{ij}}{p_{ij}} \\

\frac{p_{ij}-q_{ij}}{1-p_{ij}}  &  \frac{1-2p_{ij}+q_{ij}} {1-p_{ij}}
\end{bmatrix}
\end{equation}

Once we find the values of $\alpha_i$ and $\beta_i$, each transition matrix is independent. This implies that each link trajectory is independent evolving: it is not influenced by the behavior of the others.\\
Hence, to study the whole system's evolution, we can individually examine every transition matrix and then unify all the solutions to describe the entire system.\\
The eigenvalues and the eigenvectors of the stochastic matrices carry an interesting piece of information.

To find the eigenvalues we solve the characteristic equation: 
\begin{equation}
Det(P_{ij}-\mu^{ij} I) = \begin{vmatrix}
\frac{q_{ij}}{p_{ij}} - \mu^{ij} & \frac{p_{ij}-q_{ij}}{p_{ij}} \\

\frac{p_{ij}-q_{ij}}{1-p_{ij}}  &  \frac{1-2p_{ij}+q_{ij}} {1-p_{ij}} - \mu^{ij} 
\end{vmatrix} = 0
\end{equation}

From which we obtain:
\begin{equation}
\begin{split}
\mu^{ij}_{(1)} = 1 \\
\mu^{ij}_{(2)} = \frac{p_{ij}^2-q_{ij}}{p_{ij}^2-p_{ij}} \equiv \mu_{ij} 
\end{split}
\end{equation}

Hence, the stationary distribution, that is given by the eigenvector related to the unitary eigenvalue, is:

\begin{equation}
\pi_{ij} = (p_{ij},1-p_{ij})
\end{equation}

While the second eigenvalue can be interpreted as an indication of how fast the link converges to its stationary distribution, as remarked in the main text.

\subsection{Temporal Exponential Random Graphs as Maximum Entropy probability }

Here we show how the Temporal Exponential Random Graphs models (TERGMs), introduced by Hanneke in 2010 \cite{Hanneke}, can be seen as a maximum entropy probability distribution.
In \cite{Hanneke} is proposed a class of models to deal with time-varying graphs, making a Markovian assumption on the trajectory of the network:

\begin{eqnarray}
&\mathcal{P}(\mathbf{G_t},\mathbf{G_{t-1}},\mathbf{G_{t-2}}, \dots,\mathbf{G_{1}} | \mathbf{G_{0}} ) =& \nonumber \\  & = P(\mathbf{G}_t|\mathbf{G_{t-1}})P(\mathbf{G}_{t-1}|\mathbf{G_{t-2}}) \dots P(\mathbf{G}_1|\mathbf{G_{0}}) &
\label{eq:markov_prob}
\end{eqnarray}

In this way, he defines a nontrivial method of making a generalization of the ERGMs, admitting that $\mathcal{P}(\mathbf{G_t},\mathbf{G_{t-1}})$ has an ERGM form:

\begin{equation}
\mathcal{P}(\mathbf{G_t}|\mathbf{G_{t-1}},\vec{\theta}) = \frac{e^{\vec{\theta} C(\mathbf{G_t},\mathbf{G_{t-1}})}}{Z_t(\vec{\theta},G_{t-1})}
\label{eq:TERGMs}
\end{equation}
Where $C(\mathbf{G_t},\mathbf{G_{t-1}})$ is a vector of some properties depending on $G_t$ and $G_{t-1}$, and $\vec{\theta}$ are the associated lagrange multipliers.

Like the ERGMs, also the TERGMs can be seen as the result of a properly defined Entropy maximization problem. Let us start from a general case and then we show how to reproduce the functional form of the TERGMs.\\
In general the entropy of the trajectory reads:
\begin{equation}
\mathcal{S}(\mathcal{P}(\mathcal{G}))\equiv-\sum_{\mathcal{G}}
\mathcal{P}(\mathcal{G})
\ln \mathcal{P}(\mathcal{G})
\label{eq:entropy}
\end{equation}

Now, we want to find $P(\mathcal{G})$ that maximizes $\mathcal{S}(\mathcal{P}(\mathcal{G}))$, such that the expected value of some properties $C_i(\mathcal{G})$ is equal to the observed values $C_i^*$ and that $P(\mathcal{G})$ is normalized. We can find this probability solving an entropy maximization problem, using the method of the Lagrange multipliers:

\begin{eqnarray}
\frac{\partial}{\partial P(\mathcal{G})} \Bigg[ \mathcal{S}(\mathcal{P}(\mathcal{G})) + \alpha\left(1 - \sum_{\mathcal{G}}P(\mathcal{G}) \right) + \nonumber\\
\sum_i\theta_i \left(C_i^* - \sum_{\mathcal{G}} P(\mathcal{G}) C_i(\mathcal{G}) \right) \Bigg]
\label{etropy_maximization}    
\end{eqnarray}

The solution of this problem is given by :

\begin{equation}
    P(\mathcal{G}|\vec{\theta}) = \frac{e^{-\sum_i \theta_i C_i(\mathcal{G})}}{Z(\vec{\theta})}
    \label{TERGM_general}
\end{equation}
That represents an ensemble of trajectories, where:
\begin{equation}
    Z(\vec{\theta}) = \sum_{\mathcal{G}} e^{-\sum_i \theta_i C_i(\mathcal{G})}
\end{equation}
Is the partition function.

Now, we can note that if 

\begin{equation}
    C_i(\mathcal{G}) = \sum_t^T \frac{C_i(G_t)}{T}
    \label{c_i_t}
\end{equation}

is the sufficient statistic, eq.(\ref{TERGM_general}) becomes:

\begin{equation}
    P(\mathcal{G}|\vec{\theta}) = \prod_t \frac{e^{-\sum_i \theta_i\frac{C_i(G_t)}{T} }}{Z(\vec{\theta})} = \prod_t P(G_t|\vec{\theta})
    \label{TERGM_c_i_t}
\end{equation}

Where

\begin{equation}
    P(G_t|\vec{\theta}) =  \frac{e^{-\sum_i \theta_i\frac{C_i(G_t)}{T} }}{Z_t(\vec{\theta})}
    \label{P_gt}
\end{equation}

And 

\begin{equation}
    Z_t(\vec{\theta}) = \sum_{G_t} e^{-\sum_i \theta_i \frac{C_i(G_t)}{T}}
\end{equation}

Equation (\ref{TERGM_c_i_t}), corresponds to the scenario described in (\ref{eq:Pfactor}).\\
On the other hand, if we have

\begin{equation}
    C_i(\mathcal{G}) = \sum_t^T \frac{C_i(G_t|G_{t-1})}{T}
    \label{c_i_t-1}
\end{equation}

We obtain that

\begin{equation}
    P(\mathcal{G}|\vec{\theta}) = \prod_t P(G_t|G_{t-1},\vec{\theta})
    \label{TERGM_c_i_t-1}
\end{equation}

Where
\begin{equation}
P(G_t|G_{t-1},\vec{\theta}) =  \frac{e^{-\sum_i \theta_i\frac{C_i(G_t|G_{t-1})}{T} }}{Z_t(\vec{\theta},G_{t-1})}
\label{eq:P_gt1}
\end{equation}

with 

\begin{equation}
Z_t(\vec{\theta},G_{t-1}) = \sum_{G_t} e^{-\sum_i \theta_i \frac{C_i(G_t|G_{t-1})}{T}}
\end{equation}
Equation (\ref{eq:P_gt1}) is a particular case of the scenario indicated in equation (\ref{eq:Pnonfactor}) of the main text, and corresponds to the functional form of the TERGMs (\ref{eq:TERGMs}).

It is worth to underline that keeping the parameters time independent is like assuming that the data generating process is not changing over time: we are working with a system at the equilibrium, so the resulting model in (\ref{TERGM_c_i_t-1}) is an homogenous Markov process.\\
As we show in the main text, even if we consider the system at the equilibrium, we can still say something related to the behavior out of the equilibrium (i.e., relaxation time).

\end{document}